\documentclass[aps,prd,preprint,showpacs,showkeys,superscriptaddress,groupedaddress]{revtex4}
\usepackage{graphicx}
\usepackage{epstopdf}
\usepackage{amssymb,amsmath,bm}
\usepackage[section]{placeins}
\usepackage[font={small,it}]{caption}
\begin{large}
\end{large}
\begin{document}
\title{Effect of polarization on the structure of electromagnetic field and spatiotemporal  distribution of $e^+e^-$ pairs by colliding laser pulses}
\author{Chitradip Banerjee}
\email[Electronic address: ]{cbanerjee@rrcat.gov.in}
\affiliation{Laser Plasma Division, Raja Ramanna Centre for Advanced Technology, Indore 452013, India}
\affiliation{Homi Bhaba National Institute, Mumbai 400094, India}
\author{Manoranjan P. Singh}
\email[Electronic adress: ]{mpsingh@rrcat.gov.in}
\affiliation{Laser Plasma Division, Raja Ramanna Centre for Advanced Technology, Indore 452013, India}
\affiliation{Homi Bhaba National Institute, Mumbai 400094, India}
\begin{abstract}
Electron-positron pair production by means of vacuum polarization in the presence of strong electromagnetic (EM) field of two counterpropagating laser pulses is studied. A 3-dimensional model of the focused laser pulses based on the solution of the Maxwell's equations proposed by Narozhny and Fofanov is used to find the structure of EM field of the circularly polarized counterpropagating pulses. Analytical calculations show that the electric and magnetic fields are almost parallel to each other in the focal region when pulses are completely transverse either in electric (e-wave) or magnetic (h-wave) field. On the other hand the electric and magnetic fields are almost orthogonal when the counterpropagating pulses are made up of equal mixture of e- and h- polarized waves.  It is found that while the latter configuration of the colliding pulses has much larger threshold for pair production it can provide much shorter electron/positron pulses compared to the former case. The dependence of pair production and its spatiotemporal distribution  on polarization of the laser pulses is analyzed using the structure of the EM field.
\end{abstract}
\maketitle
\section{Introduction}
  Electron-positron ($e^-e^+$) pair production is one of the important phenomena for studying the non-linear processes in the presence of strong field interacting with the massive (fermionic) vacuum state in the realm of quantum electrodynamics (QED) \cite{RevModPhys.84.1177DiPiazza}. In QED, the vacuum is not an empty space but is a full of virtual $e^-e^+$ pairs. The word virtual means that the lifetime and the separation between electron and positron of these pair are shorter than the Compton time and length scales respectively so as to satisfy the Heisenberg uncertainty principle. If  these virtual pairs are separated up to a length scale $\lambdabar (= \hbar/m_ec)$ and over a time interval $\tau (= \hbar/m_ec^2)$, they become the real pairs.  The strength of electric field needed to have a real pair is the characteristic field of QED. It is also known as the Schwinger limit and its value is  $1.32\times 10^{16}\textnormal V/cm$ \cite{Sauter}. In the presence of such a field the vacuum becomes unstable and is depleted to $e^-e^+$ pairs. The process in which the $e^-e^+$ pairs are generated from vacuum in the presence of such constant electric field, is known as the Schwinger mechanism \cite{PhysRev.82.664}. This mechanism is very interesting because it is a non-perturbative process. From the theoretical point of view the imaginary part of the Heisenberg-Euler Lagrangian density \cite{Heisenberg} of the electromagnetic (EM) field interacting with the vacuum of the charged particles with spins $0$ and $1/2$ gives rise to particle-antiparticle pair generation which was obtained in an explicit form  in  \cite{PhysRev.82.664} wherein electric field was taken to be uniform in space and time.  The value of the uniform electric field strength is so huge that this process can not be realized experimentally because of the unavailability of such fields in laboratory.  The only way of studying such a process is by using a time-varying fields  of the ultrafast and ultraintense lasers \cite{PhysRevSTAB.5.031301}.\\
At present the available laser intensity is of the order of $\approx 10^{22}\textnormal W/cm^2$, which is still far below the critical intensity $I_{cr}=\frac{c}{4\pi}E_S^2 \approx 4.6\times 10^{29}\textnormal W/cm^2$. Several projects have been undertaken throughout the world to achieve intensities of the order of  $I\approx I_{cr}$. The SLAC (Standford Linear Accelerator Centre) \cite{PhysRevLett.79.1626SLAC} team carried out an experiment to investigate the non-linear QED processes accompanying the interaction of high energy electrons and photons with laser pulses. The experiment on  the non-linear Compton scattering of $46.6$-GeV electrons by a laser pulse with intensity of $10^{18}\textnormal W/cm^2$ was performed \cite{CbulaPhysRevLett.76.3116}.   $e^-e^+$ pairs were observed when laser photons which were backscattered up to several GeVs by the $46.6$-GeV electron beam interacted with a pulse of the second laser beam\cite{PhysRevLett.79.1626SLAC}.\\
Many theoretical studies on particle production via Schwinger mechanism have been carried out for both space and time varying fields \cite{PhysRevD.2.1191Itzykson} as given in \cite{PhysRevE.69.036408,PhysRevLett.104.220404} and the cited papers therein.
These studies have explored the pair production mechanism by using focused laser pulses having both circular and linear polarizations. It has been investigated how the pair creation mechanism depends on the electric field intensity for a single, two counter-propagating, and  multiple-colliding laser pulses \cite{PhysRevLett.104.220404}. It has been established now that it is possible to have pair creation for laser intensities much smaller than the critical intensity. These investigations have shown further that the threshold value of the intensity of the electric field of two pulses for producing a pair is about two orders less than that for a single laser pulse. The reported electric field intensity threshold value for the  two counter-propagating ultra-short laser pulse is $0.033E_S$  for laser wavelength $\lambda = 1\mu m$ and pulse duration $\tau = 10\textnormal fs$ \cite{bula}.\\

Bell et al. \cite{PhysRevLett.101.200403Bell} have demonstrated a mechanism of $e^+e^-$ pair production due to the interaction of high intensity ($\approx 10^{24}W/cm^2$) counterpropagating laser beams with the accelerated $e^-$ beam. 
In \cite{PhysRevLett.102.080402Ruf} Ruf et al. have studied a scheme of $e^+e^-$ pair production by the counterpropagating laser pulses by solving the Dirac equation numerically. They reported a characteristic modifications of the particle spectra and the Rabi oscillation dynamics. The narrow peak splitting of the resonant pair production probability served as a sensitive probe of the quasi-energy band structure. 
 Hebenstriet et al. \cite{PhysRevD.82.105026Hebenstreit} have used Dirac-Heisenberg-Wigner formalism to investigate  space-and time characteristics of nonperturbative $e^+e^-$ pair generation process by various types of electric fields. 
 The dynamics of the electron-positron pair plasma has been studied in \cite{PhysRevLett.106.035001Nerush} by Nerush et al. in a strong laser field. It has been shown that QED effects can be experimentally studied with soon-coming laser facilities like Extreme Light Infrastructure  (ELI)  and  High Power laser Energy Research (HiPER) \cite{NaturePublishingGroupDunne}. Su et al. in \cite{PhysRevA2012Su}  have investigated the particle creation process in the presence of magnetic field perpendicular to the electric field.
 It has also been shown \cite{PhysRevLett.109.253202Su} that the magnetic field can diminish the pair production. 
Gonoskov et al. \cite{PhysRevLett.111.060404Gonoskov}  have employed e-dipole field for investigating the pair production process. It has been shown that it has maximum field strength conversion efficiency compared to other focused field models (Narzhony-Fofanov \cite{Narozhny2000} and Fedotov \cite{fedotov2009electron}).
 Using the quantum kinetic theory, Kohlf\"urst et al. have studied \cite{PhysRevD.88.045028Kohlfurst} the dynamically assisted Schwinger pair-creation process  for the electric field dependent only on time.
 Spin polarized $e^+e^-$ pair production via elliptical polarized laser fields has been reported by W\"ollert et al. in \cite{PhysRevD.91.125026Wollert}. \\
In this paper we study the pair creation mechanism for different state of beam polarization of the EM field  of two counterpropagating laser pulses. The characteristic parameter for the beam polarization is defined by the parameter of asymmetry $\mu$ \cite{narozhny2005relativistic} between e- and h-waves in the field expression at any arbitrary space time position. The main aim of revisiting this topic is to know how the pairs are distributed spatially and temporally for different values of $\mu$. It has been reported that while the counterpropagating beams of entirely e- and h-polarized $(\mu = \mp 1)$ effective for pair production, the beams with equal mixture of e- and h-polarization  $(\mu = 0)$ are the worst for pair production. In this paper we analyse this observation from the structure of underlying fields. Here we restrict our attention to these three cases of polarization state only, e.g., totally e- waves $(\mu = -1)$, h-waves$(\mu = 1)$, and equal mixture of e- and h-waves $(\mu = 0)$. While $\mu = 0$ case is not suitable for efficient pair production, it is found to be appropriate for generating shorter pulses of electrons and positrons.\\

This paper is organized as follow. In the Sect. \textbf{II} we discuss the basics understanding of the pair creation mechanism in the presence of EM field. The employed field model and the modification due to finite pulse duration is also discussed in this Section. The structure of EM fields, the field invariants and the fields in the Lorentz transformed frame is analysed for different values of $\mu$, with the reference to its possible role in pair production. In Sect. \textbf{III} we discuss the spatial distribution of EM fields, as function of the normalized longitudinal and transverse coordinates,  $\chi$ and  $\xi$ respectively (defined below) and also the azimuthal angle $\phi$. The fields are compared with those in the reference frame in which the electric and magnetic fields are parallel. The polarization dependence of the temporal distribution of the pairs for different values of $\mu$ is also presented in this section. We conclude in Sect. \textbf{IV}.
\section{Theoretical Background and Field Model}
The vacuum depletion probability in the presence of constant electric and magnetic field is given by the semiclassical theory \cite{berestetskii2012quantum},
\begin{equation}
|C_V |^2=|< 0|S_{if} |0>|^2 \propto \exp(-{2 Im\mathcal{L} \textnormal{VT}}/\hbar).
\end{equation}
 Here the initial (i) and final (f) vacuum states are taken as asymptotically far away in time $(t\rightarrow\mp\infty)$ and $\textnormal{VT}$ is the 4-focal volume. $S_{if}$ is the $S$- matrix element between the initial state 'i' and the final state 'f' state interacting with a strong uniform EM field. $Im\mathcal{L}$ is the imaginary part of the Heisenberg-Euler Lagrangian density for the interaction of EM field with the vacuum state of the spin $1/2$ charged particles \cite{Heisenberg}. The vacuum depletion gives rise to $e^-e^+$ particle production. The number of pairs created per unit volume per unit time is given by the Schwinger formula \cite{PhysRev.82.664} as
 \begin{equation}
 \label{w_e_average}
 w_{e^- e^+}=\frac{e^2 E_S^2}{4 \pi^2 \hbar^2 c}  \epsilon  \eta \coth(\frac{\pi \eta}{\epsilon})\exp(-\frac{\pi}{\epsilon}).
 \end{equation}
 As the EM field associated with a typical laser pulse having wavelength of the order of a micron and pulse duration of the order of 10 fs can be taken to be uniform in space and time over the compton length and time scales,  the total number of created pairs is calculated by the integration over whole space and time. We have \cite{PhysRev.82.664}-\cite{PhysRevLett.104.220404}
\begin{equation}
 \label{N_e_average}
 N_{e^- e^+}= \int dV\int dt w_{e^- e^+} = \frac{e^2 E_S^2}{4 \pi^2 \hbar^2 c} \int dV \int dt  \epsilon  \eta \coth(\frac{\pi \eta}{\epsilon})\exp(-\frac{\pi}{\epsilon}).
\end{equation} 
Here $\epsilon = \mathcal{E}/E_S,$  $\eta = \mathcal{H}/E_S$, and $(\mathcal{E},\mathcal{H}) = \sqrt{(\mathcal{F}^2+\mathcal{G}^2)^{1/2} \pm \mathcal{F}}$ are the invariants that have the meaning of the electric and magnetic field strengths in the reference frame in which they are parallel to each other. $\mathcal{F} = \frac{1}{4}  F^{\mu \nu} F_{\mu \nu}=\frac{1}{2} (\textbf{E}^2-\textbf{H}^2 )$, $\mathcal{G} = \frac{1}{4} \epsilon ^{\mu \nu \rho \sigma} F_{\mu \nu} F_{\rho \sigma} = \textbf{E}\cdot\textbf{H}$ are Lorentz invariants of EM field. $F^{\mu\nu}$ is EM field tensor defined as $F^{\mu\nu}=\partial^{\mu}A^{\nu}-\partial^{\nu}A^{\mu}$ for EM four potential $A^{\mu}$ and $\mu,\nu$ are the Greek indices which run from $0,1,2,3$. $\epsilon ^{\mu \nu \rho \sigma}$ is totally antisymmetric Levi-Civita tensor with $\epsilon ^{0123} = 1$ \cite{Abramowitz:1974:HMF:1098650}. It is equal to $-1$ for non cyclic permutation of indices and is zero if any two indices are equal.  For a  plane wave both the Lorentz invariants $\mathcal{F}$ and $\mathcal{G}$ are zero and therefore pair production is not possible whatever maybe the intensity. To have the non-zero Lorentz invariants one can use focused EM fields. Depending on the values of $\epsilon$ and $\eta$ in the focal region, the expression of average particle creation can be approximated in a few special cases as follows:
 \begin{itemize}
 \item For $\eta\xrightarrow{} 0$ 
 \begin{equation}
 \lim\limits_{\eta\xrightarrow{}0 }\epsilon\eta\coth(\pi\eta/\epsilon) = \epsilon^2/\pi \nonumber
 \end{equation} 
 \begin{equation}
 \label{N_eta0}
 N_{e^- e^+} \approx \frac{e^2 E_S^2}{4 \pi^3 \hbar^2 c} \int dV \int dt  \epsilon^2\exp(-\frac{\pi}{\epsilon})
 \end{equation}
 \cite{PhysRevLett.104.220404}.
 \item When $0<\eta<\epsilon$, then we have $\coth(\pi\eta/\epsilon) \approx (\epsilon/{\pi\eta}+\pi\eta/{3\epsilon})$, so the approximate form of  Eq. \ref{N_e_average} as,
 \begin{equation}
 \label{N_0<eta<epsilon}
 N_{e^- e^+}\approx\frac{e^2 E_S^2}{4 \pi^2 \hbar^2 c} \int dV \int dt  (\epsilon^2/\pi+\pi\eta^2/3) \exp(-\frac{\pi}{\epsilon}).
 \end{equation}
 \end{itemize}
 We consider the solution of the 3-dimensional field model based on the Maxwell's equations proposed by Narozhny and Fofanov \cite{Narozhny2000}. According to this model the focused EM field does not possess any definite state of polarization. However it can always be represented as a superposition of e-and h-wave. Here e (h)-wave is the totally transverse electric (magnetic) field with respect to the propagation direction \cite{Narozhny2000}. For a circularly polarized laser beam propagating in the z-direction and having its focal region at the origin one can write the expression for electric and magnetic fields at any arbitrary position and time as:
 \begin{equation}
  \begin{aligned}
  \textbf{E}(\textbf{r},t) &= \frac{(1-\mu)}{2}\textbf{E}^e+\frac{(1+\mu)}{2}\textbf{E}^h, \\
  \textbf{H}(\textbf{r},t) &= \frac{(1-\mu)}{2}\textbf{H}^e+\frac{(1+\mu)}{2}\textbf{H}^h.
  \end{aligned}
  \end{equation}
  Here $\textbf{E}^e$, and $\textbf{H}^e$ are the circularly e-polarized electric and magnetic fields given as \cite{Narozhny2000}:
\begin{multline}
\label{Electric_magnetic_field_e_single_beam}
  \textbf{E}^e = iE_0e^{-i\varphi}\{F_1(\textbf{e}_x\pm i\textbf{e}_y)-F_2e^{\pm 2i\phi}(\textbf{e}_x\mp i\textbf{e}_y)\},\\
  \textbf{H}^e = \pm E_0e^{-i\varphi}\{(1-i\Delta^2\frac{\partial}{\partial\chi})[F_1(\textbf{e}_x\pm i\textbf{e}_y)+F_2e^{\pm 2i\phi}(\textbf{e}_x\mp i\textbf{e}_y)]+2i\Delta e^{\pm i\phi}\frac{\partial F_1}{\partial\xi}\textbf{e}_z\}.
\end{multline}
  Here, $\omega$ is the central frequency of the laser pulse, $\lambda$ is the wavelength, $\Delta$ is the focusing or spatial inhomogeneity parameter; $x$, $y$, and $z$ are the spatial coordinates; and 
 \begin{eqnarray}
 \varphi = \omega (t-z/c),  \xi = \rho/R,  \chi = z/L, \\
   \rho = \sqrt{x^2+y^2},  \exp(i\phi) = (x+iy)/\rho,\\
   \Delta = c/{\omega R} = \lambda/{2\pi R},  L = R/\Delta.
 \end{eqnarray} 
  In Eqn.(\ref{Electric_magnetic_field_e_single_beam}), $\pm$ sign corresponds to right and left circular polarizations respectively. We will consider only the right circular polarization for all the expressions henceforth. It should be noted, however that the all results discussed in this paper and the conclusions thereof remain the same for the left circularly polarized case too. 
 The expression for circularly polarized h-fields  can be calculated by the duality transformation of the EM field as \cite{Narozhny2000}:
 \begin{equation}
 \textbf{E}^h =  i\textbf{H}^e,      \textbf{H}^h = - i\textbf{E}^e.
 \end{equation} 
 For the weakly focused EM field i.e., $\Delta \ll 1$ the form functions $F_1$ and $F_2$ have the form of a Gaussian beam \cite{Narozhny2000};
\begin{eqnarray}
F_1 = (1+2i\chi)^{-2} ( 1-\frac{\xi^2}{1+2i\chi} ) \exp(-\frac{\xi^2}{1+2i\chi}), \\
F_2 = -\xi^2(1+2i\chi)^{-3} \exp(-\frac{\xi^2}{1+2i\chi}).
\end{eqnarray}
The finite temporal pulse width of the laser beam is accounted by the transformations \cite{Narozhny2000}:
\begin{equation}
\exp(-i\varphi) \rightarrow if^{\prime}(\varphi),\\ \exp(-i\varphi)\Delta  \rightarrow if(\varphi)\Delta \nonumber
\end{equation}
where $f(\varphi) = g(\varphi/{\omega t})\exp(-i\varphi)$, $g(0) = 1$ and $g$ should decreases very fast at the periphery of the pulse for $|\varphi| \gg \omega \tau$. We take $g(t/\tau) = \exp(-4t^2/\tau^2)$ in the focal plane $z=0$ and take $\tau = 10 fs$.
  
 For two counterpropagating laser pulses having only circularly e-polarized components i.e. $\mu = -1$, we have the following expressions for the electric and magnetic fields  
 \begin{multline}
 \label{mu = -1 E field}
 \textbf{E}^e = 2i E_0 ge^{-i\omega t}\frac{e^{-\frac{\xi^2}{{1+4\chi^2}}}}{(1+4\chi^2)}[\{\cos\left(\omega z/c-2\psi\right)-\frac{2\xi^2\sin{\phi}}{(1+4\chi^2)^{1/2}}\sin\left(\phi+\omega z/c-3\psi\right)\}\textbf{e}_x\\+i\{\cos\left(\omega z/c-2\psi\right)-\frac{2\xi^2\cos{\phi}}{(1+4\chi^2)^{1/2}}\cos(\phi+\omega z/c-3\psi)\}\textbf{e}_y],
 \end{multline}
 and
\begin{multline}
\label{mu = -1 H field}
\textbf{H}^e = 2i E_0 ge^{-i\omega t}\frac{e^{-\frac{\xi^2}{{1+4\chi^2}}}}{(1+4\chi^2)}[\{\sin\left(\omega z/c-2\psi\right)-\frac{2\xi^2\cos{\phi}}{(1+4\chi^2)^{1/2}}\sin\left(\phi+\omega z/c-3\psi\right)\}\textbf{e}_x\\+i\{\sin\left(\omega z/c-2\psi\right)-\frac{2\xi^2\sin{\phi}}{(1+4\chi^2)^{1/2}}\cos(\phi+\omega z/c-3\psi)\}\textbf{e}_y\\
-\frac{8\Delta\xi}{(1+4\chi^2)^{1/2}}(1-\frac{\xi^2}{2(1+4\chi^2)^{1/2}})\cos(\phi+\omega z/{c})\textbf{e}_z].
 \end{multline}
 Here we have neglected terms of the order of $\xi^4$ in (\ref{mu = -1 E field},\ref{mu = -1 H field}). Furthermore the term $\frac{2\chi\xi^2}{1+4\chi^2}$ has been omitted in the phase terms as it is negligible in comparison to the dominant term $\omega z/c$.
For $\mu = 1$ when the laser beams have only the circularly h-polarized components the electric and magnetic fields are the dual transformed of the e-polarized fields. The expressions of the electric and magnetic fields are:
\begin{multline}
\label{mu = 1 E field}
\textbf{E}^h = -2E_0 ge^{-i\omega t}\frac{e^{-\frac{\xi^2}{{1+4\chi^2}}}}{(1+4\chi^2)}[\{\sin\left(\omega z/c-2\psi\right)-\frac{2\xi^2\cos{\phi}}{(1+4\chi^2)^{1/2}}\sin\left(\phi+\omega z/c-3\psi\right)\}\textbf{e}_x\\+i\{\sin\left(\omega z/c-2\psi\right)-\frac{2\xi^2\sin{\phi}}{(1+4\chi^2)^{1/2}}\cos(\phi+\omega z/c-3\psi)\}\textbf{e}_y\\-\frac{8\Delta\xi}{(1+4\chi^2)^{1/2}}(1-\frac{\xi^2}{2(1+4\chi^2)^ {1/2}})\cos(\phi+\omega z/{c})\textbf{e}_z],
\end{multline}
 and
\begin{multline}
\label{mu = 1 H field}
\textbf{H}^h = 2E_0 ge^{-i\omega t}\frac{e^{-\frac{\xi^2}{{1+4\chi^2}}}}{(1+4\chi^2)}[\{\cos\left(\omega z/c-2\psi\right)-\frac{2\xi^2\sin{\phi}}{(1+4\chi^2)^{1/2}}\sin\left(\phi+\omega z/c-3\psi\right)\}\textbf{e}_x\\+i\{\cos\left(\omega z/c-2\psi\right)-\frac{2\xi^2\cos{\phi}}{(1+4\chi^2)^{1/2}}\cos(\phi+\omega z/c-3\psi)\}\textbf{e}_y]. 
\end{multline}
The electric and magnetic fields for $\mu = 0$ case are equal mixtures of e- and h-waves. 
\begin{equation}
\label{mu =0 field}
\textbf{E} = \textbf{E}^e+\textbf{E}^h = \textbf{E}^e + i \textbf{H}^e,\\
\textbf{H} = \textbf{H}^e+\textbf{H}^h = \textbf{H}^e - i \textbf{E}^e.
\end{equation}
The explicit expression of the electric field 
\begin{multline}
\label{Electric_field_mu0}
\textbf{E} = 2i E_0e^{-i\omega t}g \frac{e^{-\frac{\xi^2}{{1+4\chi^2}}}}{1+4\chi^2}\times\{[exp(i(\omega z/c-2\psi))-\frac{2i\xi^2}{(1+4\chi^2)^{1/2}}e^{-i\phi}\sin(\omega z/c-2\psi+\phi)]\textbf{e}_x\\+i[exp(i(\omega z/c-2\psi))-\\\frac{2i\xi^2}{(1+4\chi^2)^{1/2}}e^{-i\phi}\cos(\omega z/c-2\psi+\phi)]\textbf{e}_y-\frac{8\Delta\xi}{(1+4\chi^2)^{1/2}}(1-\frac{\xi^2}{2(1+4\chi^2){1/2}})\cos(\phi+\omega z/{c})\textbf{e}_z\},
\end{multline}
 and the magnetic field is $-\pi/2 $ out of phase with the electric field i.e., $\textbf{H} = exp(-i \pi/2)\textbf{E}$. Here $\exp(i\psi) = \frac{1+2i\chi}{r}$, $r=\sqrt{1+4\chi^2}$.\\
 Using Eqns.(\ref{mu = -1 E field}-\ref{mu = -1 H field}) we derive the expressions for the invariants for $\mu = -1$.
\begin{multline}
\label{F_mu-1}
\mathcal{F}(\mu = -1) = \frac{2E_0^2g^2e^{-\frac{2\xi^2}{{1+4\chi^2}}}}{(1+4\chi^2)^{2}}\times[\cos{2(\omega z/c-2\psi)}-\\\frac{2\xi^2}{(1+4\chi^2)^{1/2}}\{\cos(2\omega z/c-5\psi)+\cos{2\omega t}\cos(\psi-2\phi)\}+\mathcal{O}(\xi^4)],\\
\end{multline}
\begin{multline}
\label{G_mu-1}
\mathcal{G}(\mu = -1) = \frac{2E_0^2g^2e^{-\frac{2\xi^2}{1+4\chi^2}}}{(1+4\chi^2)^{2}}\times[\sin{2(\omega z/c-2\psi)}\\-\frac{2\xi^2}{(1+4\chi^2)^{1/2}}\{\sin(2\omega z/c-5\psi)+\cos{2\omega t}\sin(\psi-2\phi)\}+\mathcal{O}(\xi^4)]. 
\end{multline}\\
 The invariants for $\mu = 1$ can be calculated by the duality transformation of the EM field for $\mu = -1$.  We have 
 \begin{equation}
 \begin{aligned}
 \mathcal{F} (\mu = 1) = -\mathcal{F} (\mu = -1),\\
  \mathcal{G} (\mu = 1) = \mathcal{G} (\mu = -1).
 \end{aligned}
 \end{equation}
The expressions of $\mathcal{F}$ and $\mathcal{G}$ for $\mu = 0$  
\begin{multline}
\label{F_mu0}
\mathcal{F}(\mu = 0) = - \frac{8E_0^2g^2e^{-\frac{2\xi^2}{1+4\chi^2}}\xi^2}{(1+4\chi^2)^{5/2}}\left[\cos(2\omega t+2\phi-\psi)-\frac{\xi^2}{(1+4\chi^2)^{1/2}}\cos(2\omega t+2\phi)\right],
\end{multline}
 and 
\begin{multline}
\label{G_mu0}
\mathcal{G}(\mu = 0) =  \frac{8E_0^2g^2e^{-\frac{2\xi^2}{1+4\chi^2}}\xi^2}{(1+4\chi^2)^{5/2}}\left[\sin(2\omega t+2\phi-\psi)-\frac{\xi^2}{(1+4\chi^2)^{1/2}}\sin(2\omega t+2\phi)\right].
\end{multline}
At this point it may be worthwhile to compare the expressions of invariants $\mathcal{F}$ and $\mathcal{G}$ for $\mu = 0$ with those for $\mu = \mp 1$. First, the amplitude part of $\mathcal{F}$ and $\mathcal{G}$ for $\mu = 0$ has a factor of $\xi^2$ which makes it negligibly small in the focal region where $\xi \ll 1$. Away from the focal region $\xi^2$ increases but the amplitude is exponentially suppressed by the gaussian profile factor $e^{-\frac{2\xi^2}{1+4\chi^2}}$. Therefore the amplitude of $\mathcal{F}$ and $\mathcal{G}$ for $\mu = 0$ are always much smaller to those for $\mu = \mp 1$ which do not have $\xi^2$ factor.
 Second, the phase part of invariants for $\mu = \mp 1$ shows oscillatory behaviour along the propagation direction with a length scale of the order $2\pi c/\omega$ which is quite expected feature associated with the standing wave formation of counterpropagating laser beams. And this type of interference, which gets carried over to the reduced field invariants $\epsilon$ and $\eta$ (see Eqns.(\ref{epsilon_mu_1},\ref{eta_mu_1})) is the root cause effective pair production by the counterpropagating laser beams. However, this interference is absent for $\mu = 0$. In other words, we have this interference for countepropagating e-polarized beams or h-polarized beams but it is washed out when the colliding beams are made up with the equal mixture of e-and h-polarizations. This intriguing observation can be explained by analysing the expressions of EM fields of colliding pulses. \\
 The real parts of the $x$-and $y$-components of the electric field $ReE^e_x$ and $ReE^e_y$ from the Eqn.(\ref{mu = -1 E field}) for $\mu = -1$ 
 \begin{multline}
 \label{ReE_x_mu_1}
 ReE^e_x = 2E_0\sin{\omega t}g \frac{e^{-\frac{\xi^2}{{1+4\chi^2}}}}{1+4\chi^2}[\cos(\omega z/c-2\psi)-\frac{2\xi^2}{(1+4\chi^2)^{1/2}}\sin{\phi}\sin(\omega z/c-3\psi+\phi)],
 \end{multline}
 and
 \begin{multline}
 \label{ReE_y_mu_1}
 ReE^e_y = -2E_0\cos{\omega t}g \frac{e^{-\frac{\xi^2}{{1+4\chi^2}}}}{1+4\chi^2}[\cos(\omega z/c-2\psi)-\frac{2\xi^2}{(1+4\chi^2)^{1/2}}\cos{\phi}\cos(\omega z/c-3\psi)].
 \end{multline}
 The expressions of the electric field components for $\mu = -1$ in Eqns.(\ref{ReE_x_mu_1}-\ref{ReE_y_mu_1}) show oscillatory behaviour in longitudinal direction as well as in time. 
 The magnitude of the electric field 
 \begin{multline}
 \label{ReE_mu_1}
 ReE = \sqrt{{ReE^e_x}^2+{ReE^e_y}^2}
 \approx \frac{2E_0 ge^{-\frac{\xi^2}{1+4\chi^2}}}{(1+4\chi^2)}|\cos(\omega z/c-2\psi)|\\\times[1-\frac{\xi^2}{\cos(\omega z/c-2\psi)(1+4\chi^2)^{1/2}}\{\cos(\omega z/c-3\psi)+\cos{2\omega t}\cos(3\psi-\omega z/c-2\phi)\}+\mathcal{O}(\xi^4)].
 \end{multline}
 We note that the oscillatory behaviour in time has vanished while it has survived in $z$.
Similarly the $x$-and $y$-components of the real part of the magnetic field show oscillatory behaviour in $z$ and $t$: 
 \begin{multline}
 \label{ReH_x_mu_1}
 ReH^e_x = 2E_0\sin{\omega t}g \frac{e^{-\frac{\xi^2}{{1+4\chi^2}}}}{1+4\chi^2}[\sin(\omega z/c-2\psi)-\frac{2\xi^2}{(1+4\chi^2)^{1/2}}\cos{\phi}\sin(\omega z/c-3\psi+\phi)],
 \end{multline}
 \begin{multline}
 \label{ReH_y_mu_1}
 ReH^e_y = -2E_0\cos{\omega t}g \frac{e^{-\frac{\xi^2}{{1+4\chi^2}}}}{1+4\chi^2}[\sin(\omega z/c-2\psi)+\frac{2\xi^2}{(1+4\chi^2)^{1/2}}\sin{\phi}\cos(\omega z/c-3\psi+\phi)].
 \end{multline}
We neglect the $z$-component of the magnetic field as it is proportional with $\Delta (\ll 1)$ and this would only give a term proportional to $\Delta^2$ in the invariants. 
 In this approximation the real part of the magnitude of the magnetic field  
 \begin{multline}
 \label{ReH_mu_1}
 ReH = \sqrt{{ReH^e_x}^2+{ReH^e_y}^2}
 \approx \frac{2E_0 ge^{-\frac{\xi^2}{1+4\chi^2}}}{(1+4\chi^2)}|\sin(\omega z/c-2\psi)|\\\times[1-\frac{\xi^2}{\sin(\omega z/c-2\psi)(1+4\chi^2)^{1/2}}\{\sin(\omega z/c-3\psi)+\cos{2\omega t}\sin(3\psi-\omega z/c-2\phi)\}+\mathcal{O}(\xi^4)].
 \end{multline}
 The leading order term in the expressions of the electric and magnetic fields' magnitude in Eqns.(\ref{ReE_mu_1},\ref{ReH_mu_1}) are of the following form
 \begin{equation}
 \begin{aligned}
 ReE (\mu = -1)& \approx A |\cos{\theta}|,\\
 ReH (\mu = -1)& \approx A|\sin{\theta}|,
 \end{aligned}
 \end{equation}
 where $A$ is the slowly varying part of the amplitude and $\theta$ corresponds to argument of the fast varying part which basically gives rise to oscillation in the amplitude in $z$-direction,
 \begin{equation}
 \begin{aligned}
 A &= \frac{2E_0g\exp(-\frac{\xi^2}{{1+4\chi^2}})}{1+4\chi^2},\\
 \theta &= \frac{\omega z}{c}-2\varphi.\\\nonumber
 \end{aligned}
 \end{equation} 
 One can write the electric and magnetic fields components in the leading order term as:
 \begin{equation}
 \begin{aligned}
 ReE^e_x &= A\sin{\omega t}\cos\theta,\\
 ReE^e_y &= -A\cos{\omega t}\cos\theta,\\
 ReH^e_x &= A\sin{\omega t}\sin\theta,\\
 ReH^e_y &= -A\cos{\omega t}\sin\theta.
 \end{aligned}
 \end{equation}
 It is easy to see that such electric and magnetic fields will give invariants 
 \begin{equation}
 \begin{aligned}
 \mathcal{F} (\mu = -1) &\approx \frac{1}{2}A^2 cos{2\theta},\\
 \mathcal{G}(\mu = -1) &\approx \frac{1}{2}A^2\sin{2\theta}.
 \end{aligned}
 \end{equation}
 Thus oscillatory behaviour of the leading order term in fields with the same amplitude but with a phase shift of $\pi/2$ is responsible for the observed features in the expressions of invariants $\mathcal{F}$ and $\mathcal{G}$ for $\mu = -1$ discussed above.\\
For $\mu = 0$, the individual components of electric and magnetic fields show interference effects but the magnitude of the fields are independent of the oscillatory term along the propagation direction. In order to clarify this point we examine the real part of the electric and magnetic fields components and its magnitude. The $x$, $y$-components of the electric field in Eq (\ref{Electric_field_mu0}) can be written as  
  \begin{multline}
  \label{ReE_x_mu0:eqn}
  ReE_x = 2E_0g \frac{e^{-\frac{\xi^2}{{1+4\chi^2}}}}{1+4\chi^2}[\sin(\omega t-\omega z/c+2\psi)+\frac{2\xi^2}{(1+4\chi^2)^{1/2}}\sin(\omega z/c-3\psi+\phi)\cos(\omega t+\phi)],
  \end{multline}
  and
  \begin{multline}
  \label{ReE_y_mu0:eqn}
  ReE_y = -2E_0g \frac{e^{-\frac{\xi^2}{{1+4\chi^2}}}}{1+4\chi^2}[\cos(\omega t-\omega z/c+2\psi)-\frac{2\xi^2}{(1+4\chi^2)^{1/2}}\cos(\omega z/c-3\psi+\phi)\cos(\omega t+\phi)].
  \end{multline}
  The components consist of interference terms in longitudinal axis and in time. Using these components we calculate the magnitude of the electric field. 
 
  \begin{multline}
  \label{ReE_mu0:eqn}
  ReE = \sqrt{{ReE_x}^2+{ReE_y}^2}\\
   = 2E_0g \frac{e^{-\frac{\xi^2}{{1+4\chi^2}}}}{1+4\chi^2}[1-\frac{4\xi^2}{(1+4\chi^2)^{1/2}}\cos(\omega t+\phi)\cos(\omega t+\phi-\psi)+\frac{4\xi^4}{(1+4\chi^2)}\cos^2(\omega t+\phi)]^{1/2}.
  \end{multline}
 It is easy to see that the leading order term for the magnitude of the total electric field does not show the oscillatory behaviour of the electric field components along the z-axis. It is due to $\pi/2$ phase difference between the oscillations of the components.  The same conclusion holds good for the components and the magnitude of  the total magnetic field. For the sake of completeness  we write down  the $x$-and $y$-components of the magnetic field using Eq. (\ref{mu =0 field})
  \begin{multline}
  \label{ReH_x_mu0:eqn}
  ReH_x = 2E_0g \frac{e^{-\frac{\xi^2}{{1+4\chi^2}}}}{1+4\chi^2}[\cos(\omega t-\omega z/c+2\psi)-\frac{2\xi^2}{(1+4\chi^2)^{1/2}}\sin(\omega z/c-3\psi+\phi)\sin(\omega t+\phi)],
  \end{multline}
  and
  \begin{multline}
  \label{ReH_y_mu0:eqn}
  ReH_y = 2E_0g \frac{e^{-\frac{\xi^2}{{1+4\chi^2}}}}{1+4\chi^2}[\sin(\omega t-\omega z/c+2\psi)-\frac{2\xi^2}{(1+4\chi^2)^{1/2}}\cos(\omega z/c-3\psi+\phi)\sin(\omega t+\phi)].
  \end{multline}
Finally we calculate the magnitude of the real part of the magnetic field
  \begin{multline}
  \label{ReH_mu0:eqn}
  ReH = \sqrt{{ReH_x}^2+{ReH_y}^2}
   = 2E_0g \frac{e^{-\frac{\xi^2}{{1+4\chi^2}}}}{1+4\chi^2}[1-\frac{4\xi^2}{(1+4\chi^2)^{1/2}}\sin(\omega t+\phi)\sin(\omega t+\phi-\psi)\\+\frac{4\xi^4}{(1+4\chi^2)}\sin^2(\omega t+\phi)]^{1/2}.
  \end{multline}
The expressions of the real part of the electric and magnetic fields in Eqns.(\ref{ReE_mu0:eqn}-\ref{ReH_mu0:eqn}) show that their leading order terms  are the same. Hence in the expression for $\mathcal{F}$ which is basically the half of the difference between square of real part of the electric and magnetic fields, this term cancels out and the leading order term is proportional to $\xi^2$. Initially it increases with the increase in $\xi$ but eventually its magnitude falls off  because of the gaussian pulse profile factor. Eqns.(\ref{F_mu0}-\ref{G_mu0}) show the expression of invariants of EM fields with an equal mixture of e and h waves. These do not show oscillatory behaviour along the $z-$axis. However, the  oscillations are present behaviour present in $t$ and  $\phi$.\\
The expressions of the reduced field invariants for $\mu = -1, 1, 0$ are given as:
\begin{multline}
\label{epsilon_mu_1}
\epsilon (\mu = -1) \approx \frac{2E_0 ge^{-\frac{\xi^2}{1+4\chi^2}}}{(1+4\chi^2)}|\cos(\omega z/c-2\psi)|\times[1-\frac{\xi^2}{\cos(\omega z/c-2\psi)(1+4\chi^2)^{1/2}}\times\\\{\cos(\omega z/c-3\psi)+\cos{2\omega t}\cos(3\psi-\omega z/c-2\phi)\}+\mathcal{O}(\xi^4)]
\end{multline}
 and
\begin{multline}
\label{eta_mu_1}
\eta (\mu = -1) \approx \frac{2E_0 ge^{-\frac{\xi^2}{1+4\chi^2}}}{(1+4\chi^2)}|\sin(\omega z/c-2\psi)|\times[1-\frac{\xi^2}{\sin(\omega z/c-2\psi)(1+4\chi^2)^{1/2}}\times\\\{\sin(\omega z/c-3\psi)+\cos{2\omega t}\sin(3\psi-\omega z/c-2\phi)\}+\mathcal{O}(\xi^4)].
\end{multline}
The leading order term in the reduced fields $\epsilon$ and $\eta$ in the above Eqns.(\ref{epsilon_mu_1}-\ref{eta_mu_1}) can be approximated as,
\begin{equation}
\begin{aligned}
\epsilon & \approx A|\cos\theta|,\\
\eta & \approx A|\sin\theta|,
\end{aligned}
\end{equation}
where $A$ and $\theta$ have been defined earlier. These approximate expressions are identical to those for  the electric and magnetic fields for small values of $\chi$ and  $\xi$. In the next section we will comeback to this interesting observation and  discuss in details its possible ramification.  
For h-waves the reduced electric and magnetic fields are calculated by the duality transformation. The expressions are 
\begin{equation}
\begin{aligned}
\epsilon (\mu = 1) = \eta (\mu = -1),\\
\eta (\mu = 1) = \epsilon (\mu = -1).
\end{aligned}
\end{equation}
 The reduced fields for $\mu = 0$ are given as:
\begin{multline}
\label{eqn:epsilon_mu0}
\epsilon (\mu = 0) \approx \frac{4E_0 ge^{-\frac{\xi^2}{1+4\chi^2}}\xi}{(1+4\chi^2)^{5/4}}|\sin(\psi/2-\phi-\omega t)|\times[1-\frac{\xi^2}{4\sin^2(\psi/2-\phi-\omega t)(1+4\chi^2)^{1/2}}\{\cos(\psi-4\phi-4\omega t)\\-\cos{2(\omega t+\phi)}\}+\mathcal{O}(\xi^4)],
\end{multline}
 and 
\begin{multline}
\label{eta_mu0}
\eta (\mu = 0) \approx \frac{4E_0 ge^{-\frac{\xi^2}{1+4\chi^2}}\xi}{(1+4\chi^2)^{5/4}}|\cos(\psi/2-\phi-\omega t)|\times[1-\frac{\xi^2}{4\cos^2(\psi/2-\phi-\omega t)(1+4\chi^2)^{1/2}}\{\cos(\psi-4\phi-4\omega t)\\+\cos{2(\omega t+\phi)}\}+\mathcal{O}(\xi^4)].
\end{multline}
 The above expressions for $\epsilon$ and $\eta$ are derived in  the small $\chi$, $\xi$ approximation in order to understand the physical origin of the pair production in terms of the structure of EM fields and the invariants  in the focal region. 
It is clear that the qualitative features of the invariants $\mathcal{F}$ and $\mathcal{G}$ get translated into reduced field invariants $\epsilon$ and $\eta$. As before the amplitudes of $\epsilon$ and $\eta$ are the same in all the cases. While it is maximum for $\xi = 0$ for $\mu = \mp 1$ it identically vanishes for $\mu = 0$ case, and is much smaller for any other value of $\xi$ because of the presence of the factor $\xi$ in the leading order term for the amplitude in the latter case. For $\mu = \mp 1$ both $\epsilon$ and $\eta$ show oscillatory behaviour with phase difference of $\pi/2$ with spatial frequency $\approx 2\pi\omega/c$ in $z$-direction, the propagation direction. The origin of this oscillation, as discussed earlier, is the interference of the counterpropagating beams. This type of oscillation is absent in $\epsilon$ and $\eta$ for $\mu = 0$. However they show oscillatory behaviour in the temporal domain and with the azimuthal variable $\phi$.
We note here that in going from $\mu = -1$ to $1$, $\epsilon$ and $\eta$ get interchanged. For $\mu = -1$, $\epsilon$ shows a maximum for $\xi = 0$ and $\chi = 0$. Consequently the spatial distribution of $e^+e^-$ pairs would show a peak at the centre of the focal spot. However for $\mu = 1$ $\epsilon$ is maximum for $\xi = 0$, $\chi = 0$ and hence the spatial distribution of $e^+e^-$ pairs will show a dip right at the centre of the focal spot. We will return this point later.
Having discussed the expressions for the fields, invariants for various polarization state of counterpropagating pulses we present results in the next section.
\section{Results and Discussion}
As discussed in the previous section the structure of the electromagnetic fields and their relationship to the reduced invariant fields are quite sensitive to the polarization of the colliding pulses. We, therefore, first present the spatial variation of EM fields and the corresponding reduced fields $\epsilon$ and $\eta$ for $\mu = \mp 1$.

\subsection{Fields and particles for $\mu = \mp 1$ beams: }
\begin{figure}[h]
 \begin{center}
 \includegraphics[width=75mm]{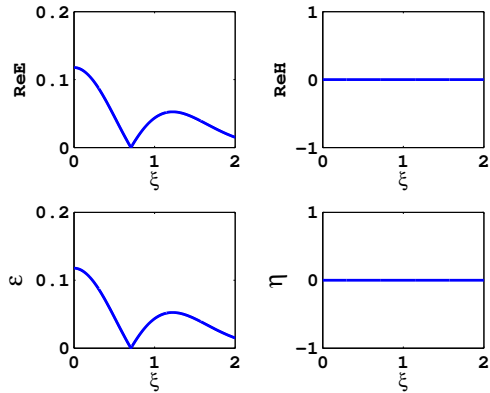}
 \end{center}
 \begin{center}
 \caption{Variation of $|Re\textbf{E}|$, $|Re\textbf{H}|$, $\epsilon$, and $\eta$ with dimensionless transverse variable $\xi$ of the two counterpropagating focused laser beam for $\mu = -1$. Fields are normalized with $E_S$, $E_0 = 0.0565$, $\chi = 0$, $\phi = 0$, and $t=0$.}\label{fig:E_H_epsilon_eta_xi_mu_1_chi0}
 \end{center}
 \end{figure}

\begin{figure}[h]
 \begin{center}
 \includegraphics[width=75mm]{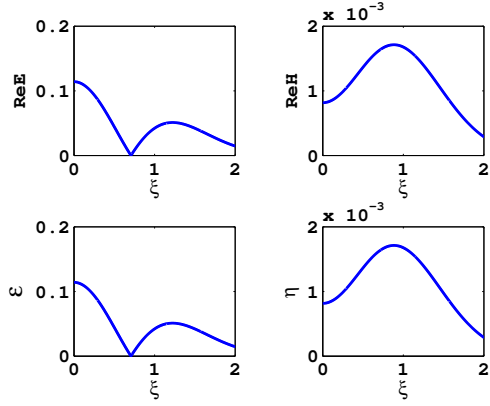}
\end{center}
\begin{center}
\caption{Variation of $|Re\textbf{E}|$, $|Re\textbf{H}|$, $\epsilon$, and $\eta$ with dimensionless transverse variable $\xi$ of the two counterpropagating focused laser beam for $\mu = 1$. Fields are normalized with $E_S$, $E_0 = 0.0565$, $\chi = 0.0164$, $\phi = 0$ and $t=0$.}\label{fig:E_H_epsilon_eta_xi_mu1_chi0164}
\end{center}
\end{figure}

\begin{figure}[h]
\begin{center}
\includegraphics[width=75mm]{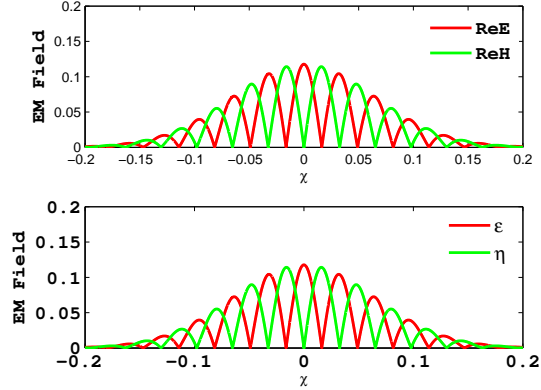}
\end{center}
\begin{center}
\caption{Variation of $|Re\textbf{E}|$, $|Re\textbf{H}|$, $\epsilon$, and $\eta$ with dimensionless longitudinal variable $\chi$ of the two counterpropagating focused laser beam for $\mu = -1$. Fields are normalized with $E_S$, $E_0 = 0.0565$, $\xi = 0$, and $t=0$.}\label{fig:E_H_epsilon_eta_chi_mu_1_2May2016}
\end{center}
\end{figure}

\begin{figure}[h]
\begin{center}
\includegraphics[width=75mm]{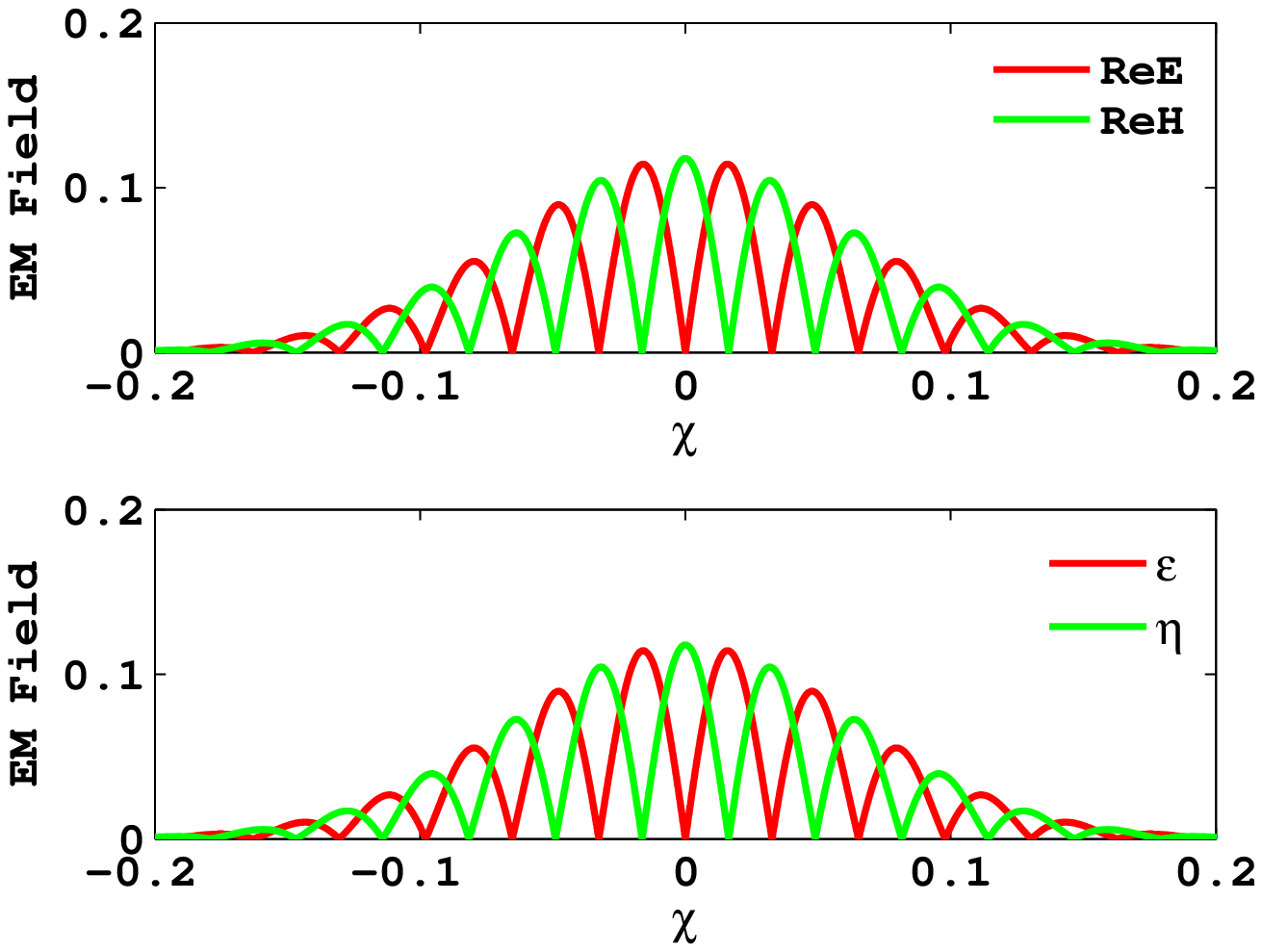}
\end{center}
\begin{center}
\caption{Variation of $|Re\textbf{E}|$, $|Re\textbf{H}|$, $\epsilon$, and $\eta$ with dimensionless longitudinal variable $\chi$ of the two counterpropagating focused laser beam for $\mu = 1$. Fields are normalized with $E_S$, $E_0 = 0.0565$, $\xi = 0$, and $t=0$.}\label{fig:E_H_epsilon_eta_chi_mu1_2May2016}
\end{center}
\end{figure}

\begin{figure}[h]
  \begin{center}
  \includegraphics[width=75mm]{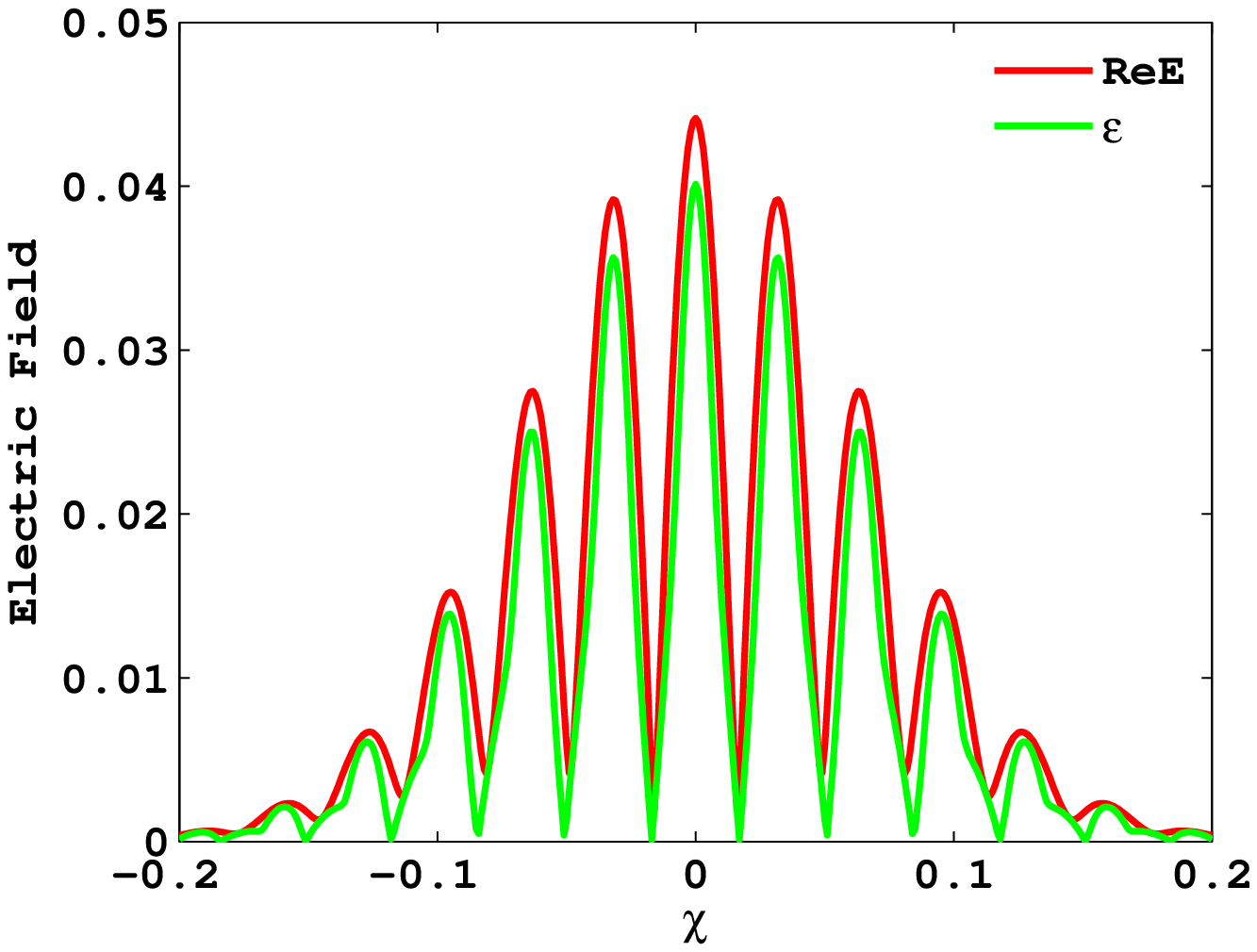}
  \end{center}
  \begin{center}
  \caption{Variation of $|Re\textbf{E}|$, and $\epsilon$ with dimensionless longitudinal variable $\chi$ of the two counterpropagating focused laser beam for $\mu = -1$. Fields are normalized with $E_S$, $E_0 = 0.0565$,  $\xi = 0.8$, $\phi = 0$ and $t=0.4 fs$.}\label{fig:non_parallel_electric_mu_1}
  \end{center}
  \end{figure}

\begin{figure}[h]
  \begin{center}
  \includegraphics[width=75mm]{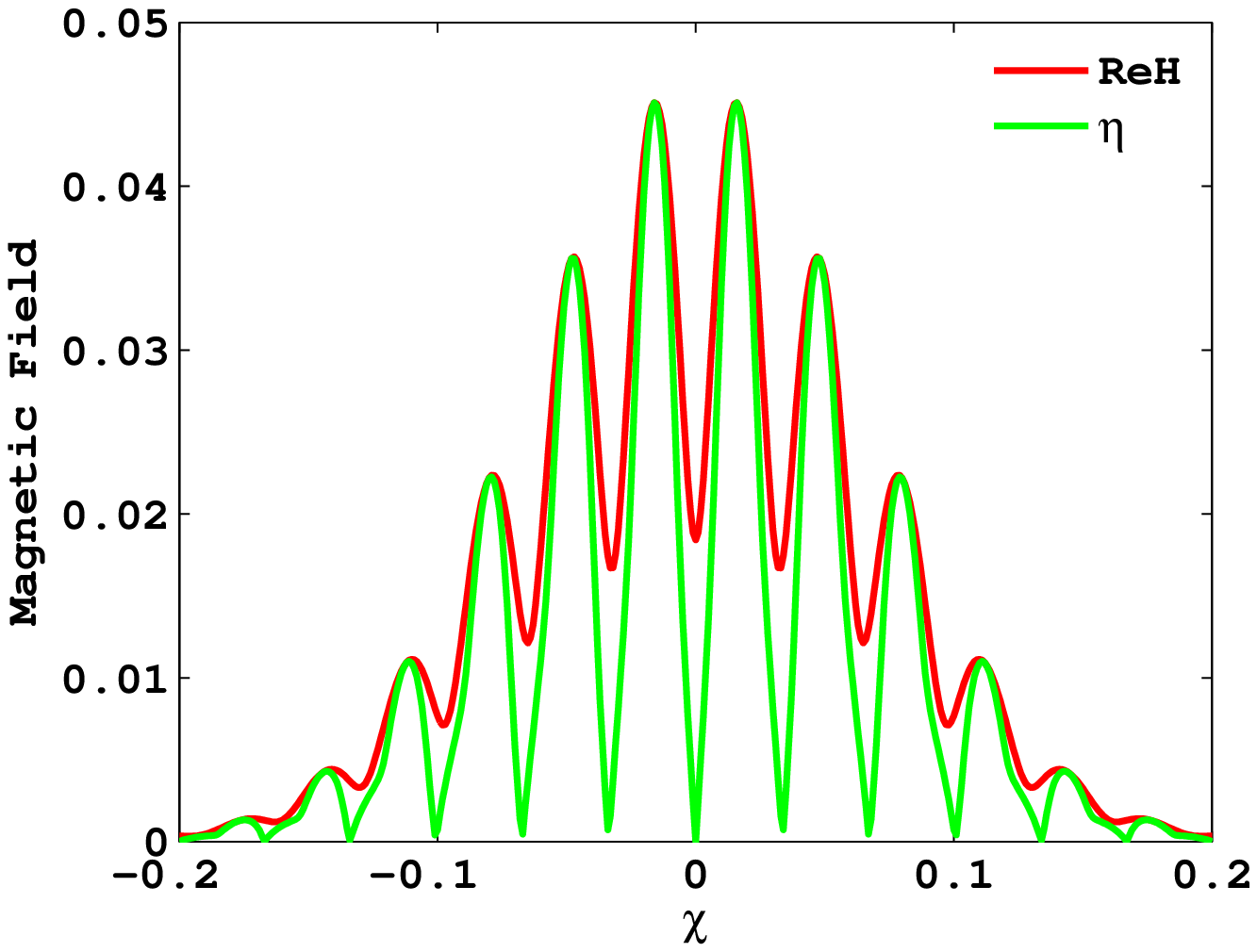}
  \end{center}
  \begin{center}
  \caption{Variation of $|Re\textbf{H}|$, and $\eta$ with dimensionless longitudinal variable $\chi$ of the two counterpropagating focused laser beam for $\mu = -1$. Fields are normalized with $E_S$, $E_0 = 0.0565$,  $\xi = 0.8$, $\phi = 0$ and $t=0.4 fs$.}\label{fig:non_parallel_magnetic_mu_1}
  \end{center}
  \end{figure}

We consider the EM field by the superposition of two counterpropagating laser pulses. For the field distribution in the focal plane we present the results from Eq.(\ref{mu = -1 E field}-\ref{mu =0 field}). Figure \ref{fig:E_H_epsilon_eta_xi_mu_1_chi0} shows the fields $|Re{\textbf{E}}|$, $|Re{\textbf{H}}|$, and invariants  $\epsilon$, and $\eta$ as a function of  $\xi$ for $\mu = -1$ at $z = 0$ plane and time $t = 0$. The electric field and $\epsilon$ show maximum at $\xi = 0$ and falls off in the peripheral region. The magnetic field and $\eta$ vanish for all values of  $\xi$ for $z = 0$-plane and at $t = 0$. 
 Figure \ref{fig:E_H_epsilon_eta_xi_mu1_chi0164} presents the fields $|Re{\textbf{E}}|$, $|Re{\textbf{H}}|$, and invariants $\epsilon$, and $\eta$ as a function of  $\xi$ for $\mu = 1$ for $z = 0.0164L$ (because electric field has one of its maxima at this value of $z$, see below) and time $t = 0$.  The variation in  $|Re\textbf{E}|$ and $\epsilon$ with $\xi$ is same as that for $\mu = -1$ case. The variation in $|Re\textbf{H}|$ and $\eta$ is somewhat different. Both of them show a maximum in the peripheral region. However, their values are always much smaller than that of $|Re{\textbf{E}}|$ or $\epsilon$. \\
 Figure \ref{fig:E_H_epsilon_eta_chi_mu_1_2May2016} shows the distribution of the fields and the reduced invariant fields as a function of normalized longitudinal coordinate $\chi$ for $\mu = -1$. Here the  field distributions form standing wave patterns because of the superposition of two monochromatic EM waves. The decrease in the amplitude of oscillation is characterized by the form function $g$ and there are  multiple maxima although the central maximum is located at $\chi = 0$. Figure \ref{fig:E_H_epsilon_eta_chi_mu1_2May2016} shows the distribution of the fields and invariant fields with normalized longitudinal coordinate $\chi$ for $\mu = 1$. Here the fields and invariants show standing wave pattern similar to that in the case of $\mu = -1$.
The longitudinal extent of the focused field in both the cases is upto $\chi = \pm 0.2$ and is symmetrical about $\chi = 0$. In this region the field distribution shows oscillation with decreasing amplitude. The maxima of these oscillations are spaced by $\approx 0.03272L$ (Rayleigh lengths) along the propagation direction . For $\mu = -1$, the central lobe shows the maximum for the electric field and the minimum for magnetic field and vice versa for $\mu = 1$. \\
The distributions of the fields as shown in Figs.(\ref{fig:E_H_epsilon_eta_xi_mu_1_chi0},\ref{fig:E_H_epsilon_eta_xi_mu1_chi0164},\ref{fig:E_H_epsilon_eta_chi_mu_1_2May2016},\ref{fig:E_H_epsilon_eta_chi_mu1_2May2016}) reveal remarkable equality between $|Re\textbf{E}| (|Re\textbf{H}|)$, and $\epsilon (\eta)$. Infact, in Figs.(\ref{fig:E_H_epsilon_eta_chi_mu_1_2May2016}-\ref{fig:E_H_epsilon_eta_chi_mu1_2May2016}), they are just identical. Recalling that $\epsilon (\eta)$ has the meaning of a transformed electric field (magnetic field) in the Lorentz frame in which the electric and magnetic fields are parallel to each other. Such a frame is achieved for any non-orthogonal electric $\textbf{E}$ and magnetic $\textbf{H}$ fields by the Lorentz boost operation given by 
\begin{equation} 
\frac{\frac{\textbf{V}}{c}}{1+\frac{V^2}{c^2}} = \frac{\textbf{E}\times\textbf{H}}{|\textbf{E}|^2+|\textbf{H}|^2},
\end{equation} 
see in Ref. {\cite{Landaufield}-\cite{griffiths1999introduction}}.\\
 The observation that in both cases ($\mu = \mp 1$) the fields in the lab frame and the reduced fields ($\epsilon$, and $\eta$) in transformed frame are identical or nearly identical suggests that the fields are parallel or nearly parallel in both the frame. This can be further understood by evaluating the cross product $\textbf{C}( = Re\textbf{E}\times Re\textbf{H})$ to calculate $\textbf{V}$. In Cartesian coordinate system 
 \begin{multline}
 C_x = ReE^e_yReH^e_z-ReE^e_zReE^e_y\\
 = -\frac{8E_0^2g^2e^{-\frac{2\xi^2}{1+4\chi^2}}\Delta\xi\sin{2\omega t}}{(1+4\chi^2)^{4}}\times[\cos(\omega z/c-2\psi)-\frac{2\xi^2}{(1+4\chi^2)^{1/2}}\cos{\phi}\cos(\omega z/c+\phi-3\psi)]\\\times[-2(1+4\chi^2)^{3/2}\cos(\omega z/c+\phi-3\psi)+\xi^2(1+4\chi^2)\cos(\omega z/c+\phi-4\psi)],
 \end{multline} 
 \begin{equation}
 V_x \approx -\frac{\Delta\xi}{(1+4\chi^2)^{2}}\sin{2\omega t}\cos(\omega z/c-2\varphi),
 \end{equation}
 \begin{multline}
 C_y = ReE^e_zReH^e_x-ReE^e_xReH^e_z\\
 = -\frac{8E_0^2g^2e^{-\frac{2\xi^2}{1+4\chi^2}}\Delta\xi(1-cos{2\omega t})}{(1+4\chi^2)^{4}}[\cos(\omega z/c-2\psi)-\frac{2\xi^2}{(1+4\chi^2)^{1/2}}\sin{\phi}\sin(\omega z/c+\phi-3\psi)]\\\times[-2(1+4\chi^2)^{3/2}\cos(\omega z/c+\phi-3\psi)+\xi^2(1+4\chi^2)\cos(\omega z/c+\phi-4\psi)],
 \end{multline}
 \begin{equation}
  V_y \approx -\frac{2\Delta\xi}{(1+4\chi^2)^{2}}\sin^2{\omega t}\cos(\omega z/c-2\varphi),
  \end{equation}
 and 
 \begin{multline}
 C_z = ReE^e_xReH^e_y-ReE^e_yReH^e_x\\
 = -\frac{4E_0^2g^2e^{-\frac{2\xi^2}{1+4\chi^2}}\xi^2\sin{2\omega t}}{(1+4\chi^2)^{5/2}}[\sin(2\phi-\psi)-\frac{\xi^2}{(1+4\chi^2)^{1/2}}\sin{2(\omega z/c+\phi-3\psi)}],
 \end{multline}
 \begin{equation}
  V_z \approx -\frac{\xi^2}{2(1+4\chi^2)^{1/2}}\sin{2\omega t}\sin(2\phi-\varphi).
  \end{equation} 
 As both $\xi$ and $\Delta\ll 1$, $\textbf{V}$ is negligibly small in the focal region and vanishes for $\xi = 0$. This explains the observation that the transformation from ($Re\textbf{E}$, $Re\textbf{H}$) to ($\epsilon$, $\eta$) is nearly identity transformation in the focal region and for the special case of $\xi = 0$ it is exactly identity transformation. The physical consequence of a very small value of $|\textbf{C}|$, in the focal region is that a very small amount of EM energy is flows out of the focal region and thereby resulting in an efficient pair production for two beam configuration for $\mu = \pm 1$. Furthermore, since $|\textbf{C}|$ is proportional to $\Delta$ a smaller value of $\Delta$ will lead to a larger number of pairs. This effect has been attributed to the increase in the focal volume in the literature \cite{bula}. However, the explanation given here is more direct and physical.\\  
 The components of $Re\textbf{E}\times Re\textbf{H}$ are proportional to
 $\xi$. Therefore,  the electric and magnetic fields are not completely parallel as one
 moves to the regions in the focal volume where $\xi \ne 0$. This leads a
 small but finite amount of energy to flow out of the focal region due to
 the slight non-parallelism of the fields.
  In what follows we examine the effect of this on the possible mismatch
 between the electric and magnetic fields in both the frames. We show
 plots of  $(\left|Re\textbf{E}\right|,\epsilon)$ and $\left(|Re\textbf{H}|,
 \eta\right)$ in 
 Figs.(\ref{fig:non_parallel_electric_mu_1},\ref{fig:non_parallel_magnetic_mu_1})
 respectively as a function of the scaled longitudinal variable $\chi$ for
 the values of $\xi=0.8$ and $t=0.4 fs$. It is clear from these plots that
 even going away from the focal region makes very little difference in the
 fields in  both the frames.
 
\begin{figure}[h]
\begin{center}
\includegraphics[width=75mm]{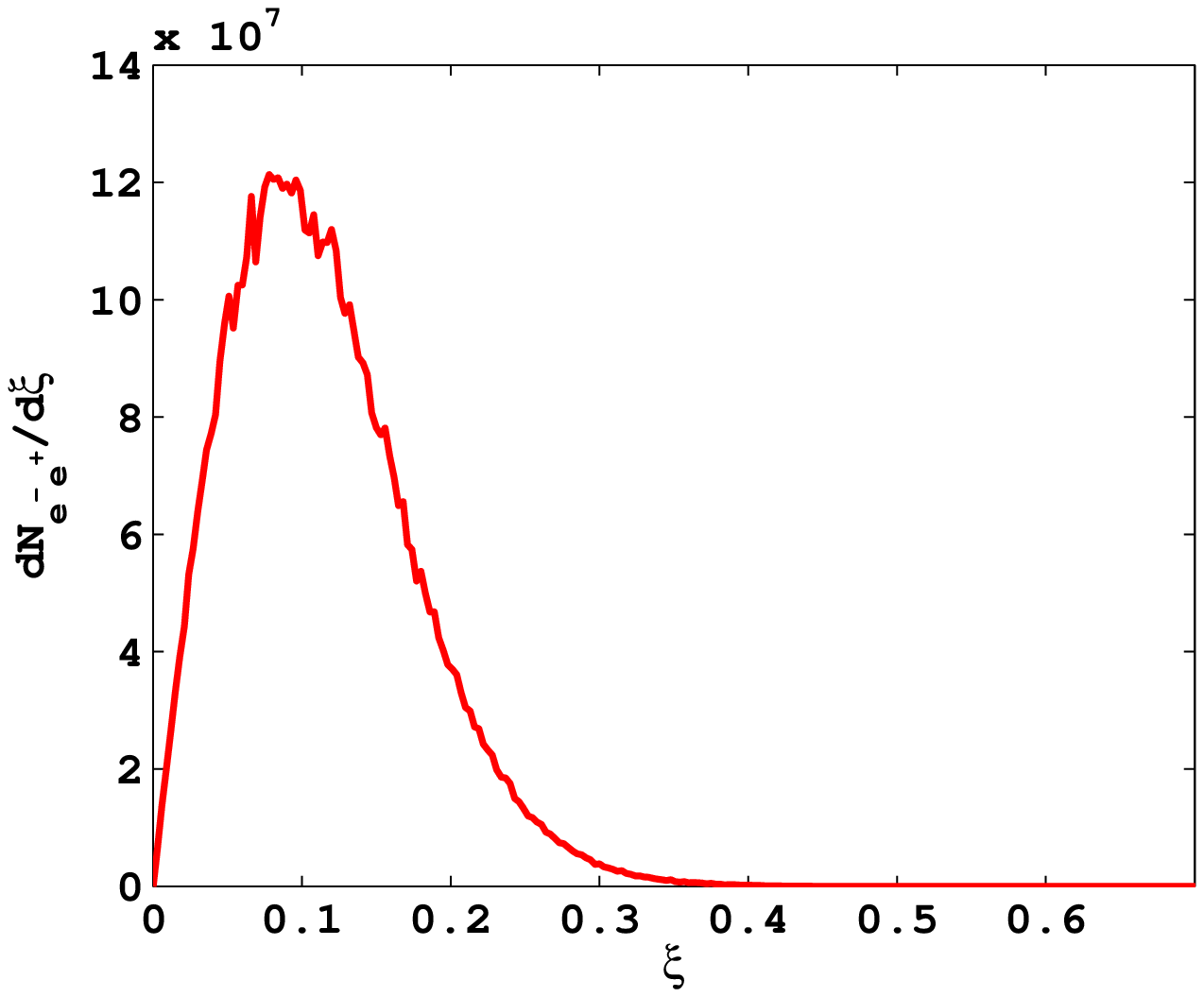}
\end{center}
\begin{center}
\caption{ Distribution of $dN_{e^+e^-}/{d\xi}$ as a function of  $\xi$ for  two counterpropagating focused laser beams with $\mu = -1$.  $E_0 = 0.0565E_S$, $\Delta = 0.1$, and $\tau=10fs$.}\label{fig:dN_vs_xi_mu_1_7May2016}
\end{center}
\end{figure}

\begin{figure}[h]
\begin{center}
\includegraphics[width=75mm]{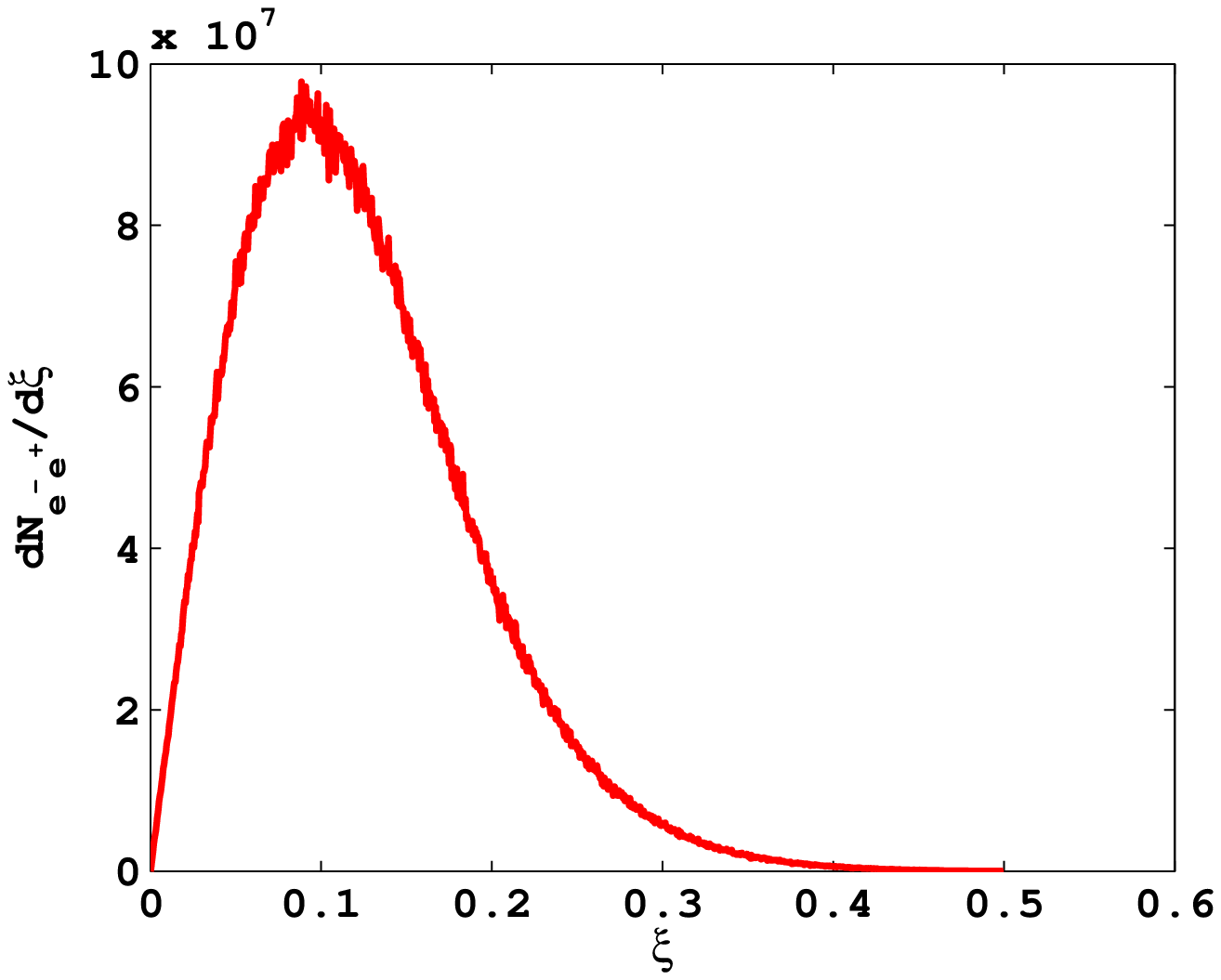}
\end{center}
\begin{center}
\caption{Distribution of $dN_{e^+e^-}/{d\xi}$ as function of  $\xi$ for two counterpropagating focused laser beams with $\mu = 1$. $E_0 = 0.0565E_S$, $\Delta = 0.1$, and $\tau=10fs$.}\label{fig:dN_vs_xi_mu1_7May2016}
\end{center}
\end{figure}

\begin{figure}[h]
\begin{center}
\includegraphics[width=75mm]{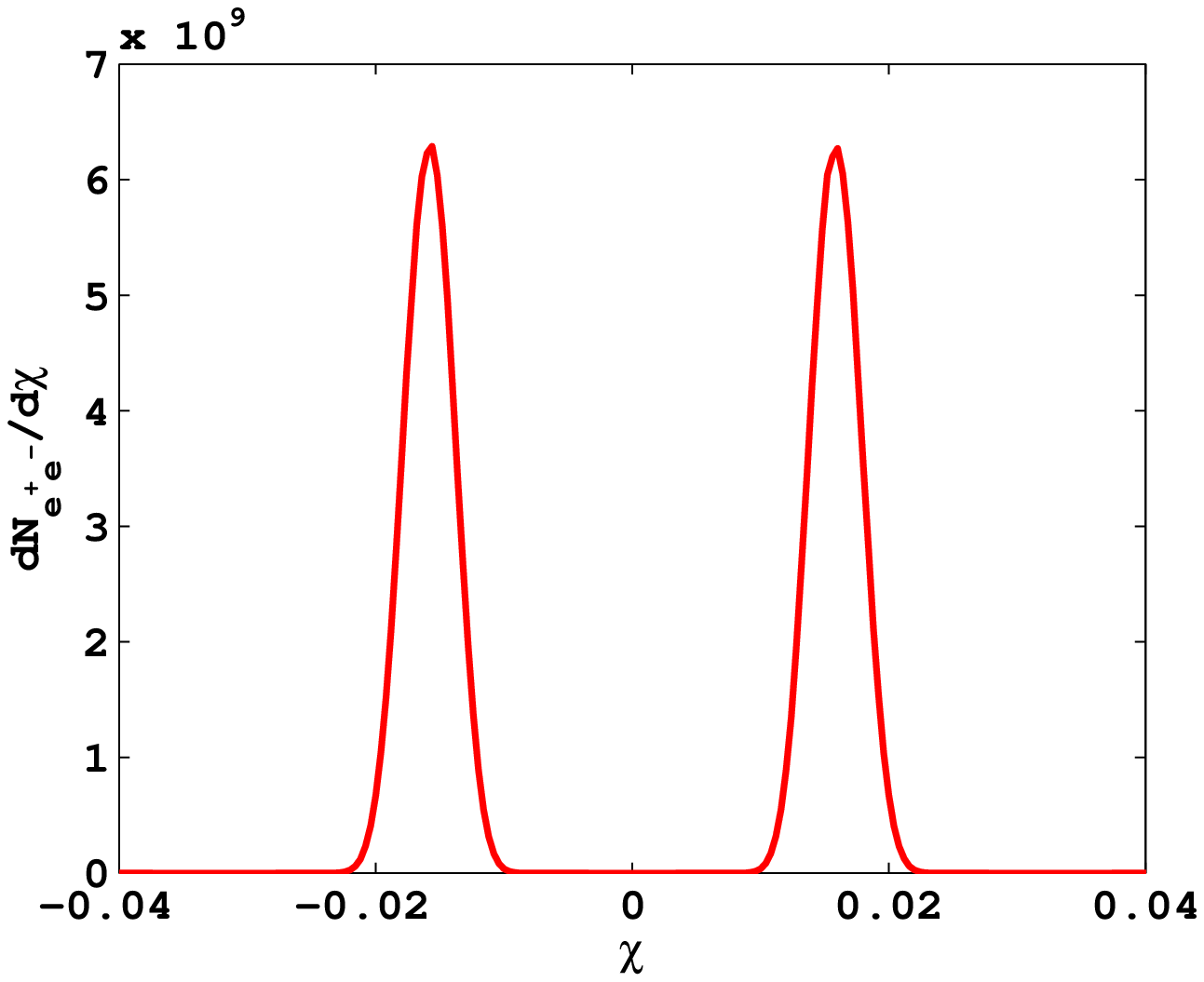}
\end{center}
\begin{center}
\caption{Distribution of $dN_{e^+e^-}/{d\chi}$ as function of $\chi$ for two counterpropagating focused laser beams having $\mu = 1$.  $E_0 = 0.0565E_S$, $\Delta = 0.1$, and $\tau=10fs$.}\label{fig:dN_chi_mu1_9May2016}
\end{center}
\end{figure}

\begin{figure}[h]
\begin{center}
\includegraphics[width=75mm]{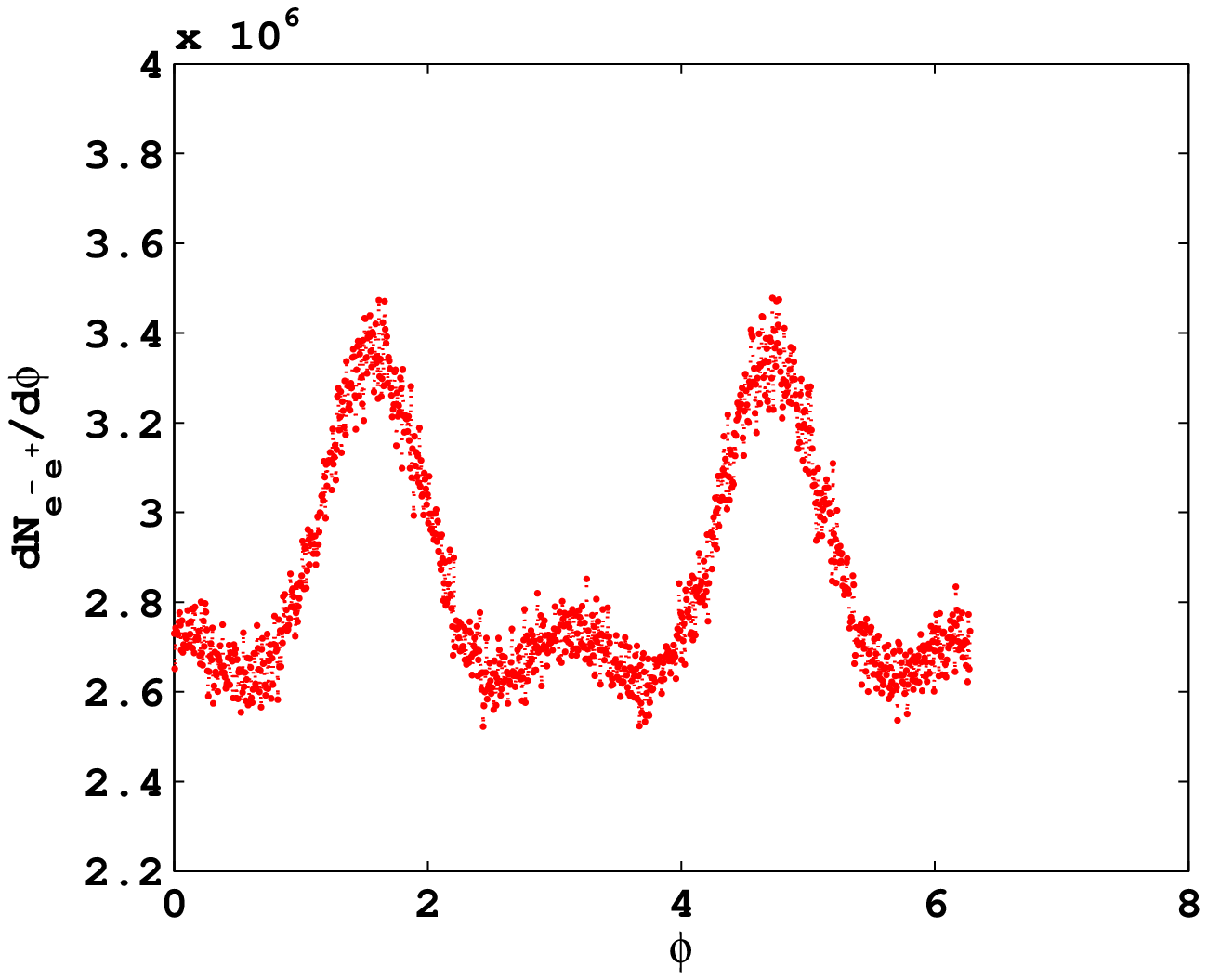}
\end{center}
\begin{center}
\caption{Distribution of $dN_{e^+e^-}/{d\phi}$ as a function of $\phi$ for two counterpropagating focused laser beam with $\mu = -1$.  $E_0 = 0.0565E_S$, $\Delta = 0.1$, and $\tau=10fs$.}\label{fig:dN_vs_phi_mu_1_7May2016}
\end{center}
\end{figure}

\begin{figure}[h]
\begin{center}
\includegraphics[width=75mm]{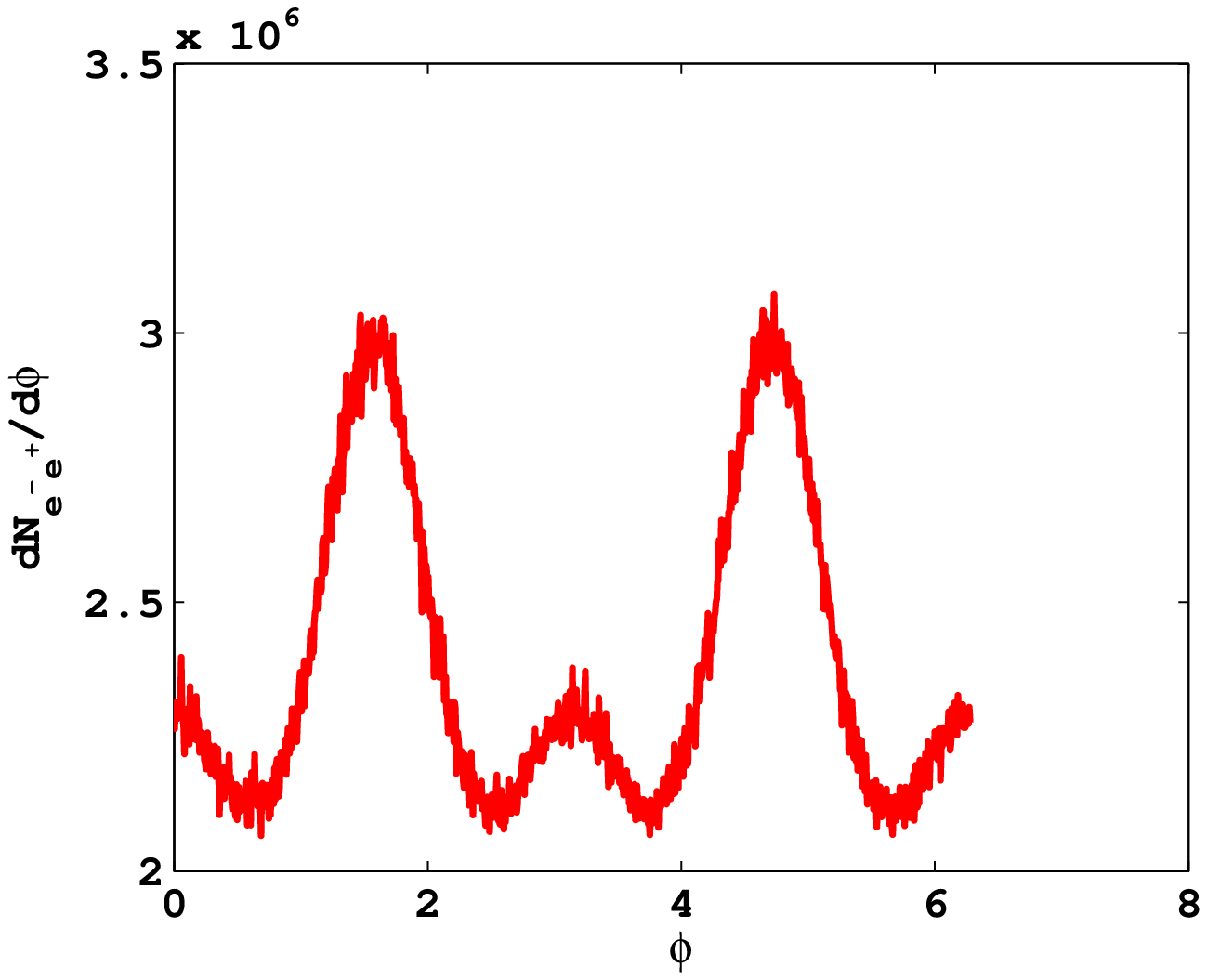}
\end{center}
\begin{center}
\caption{Distribution of  $dN_{e^+e^-}/{d\phi}$ as a function of $\phi$ for two counterpropagating focused laser beam with $\mu = 1$. $E_0 = 0.0565E_S$, $\Delta = 0.1$, and $\tau=10fs$.}\label{fig:dN_vs_phi_mu1_7May2016}
\end{center}
\end{figure}

\begin{figure}[h]
\begin{center}
\includegraphics[width=75mm]{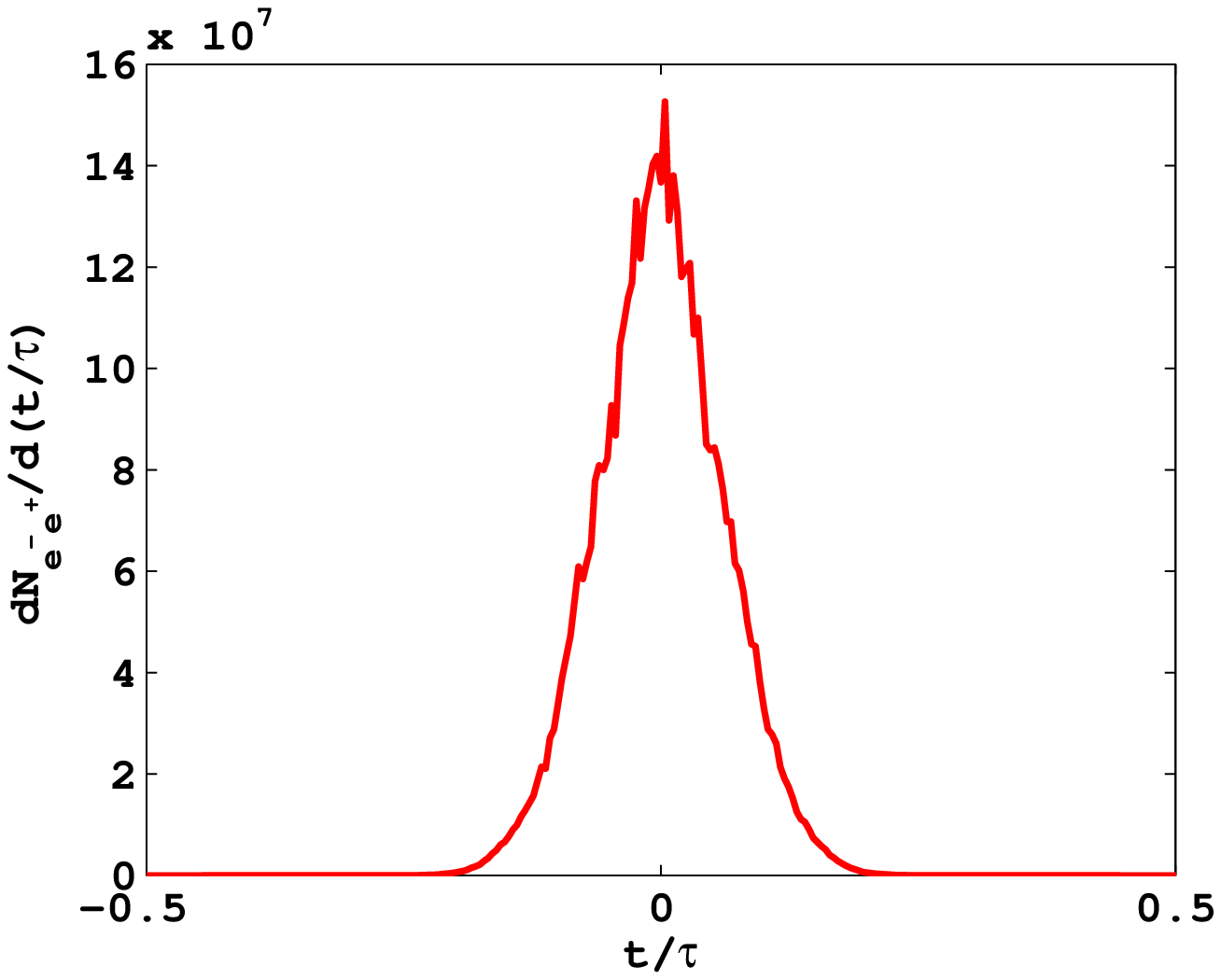}
\end{center}
\begin{center}
\caption{Distribution of $dN_{e^+e^-}/d{(t/\tau)}$ as a function of  $t/\tau$ for  two counterpropagating focused laser beams with $\mu = -1$.  $E_0 = 0.0565E_S$, $\Delta = 0.1$, and $\tau=10fs$.}\label{fig:dN_vs_t_mu_1_8May2016}
\end{center}
\end{figure}

\begin{figure}[h]
 \begin{center}
 \includegraphics[width=75mm]{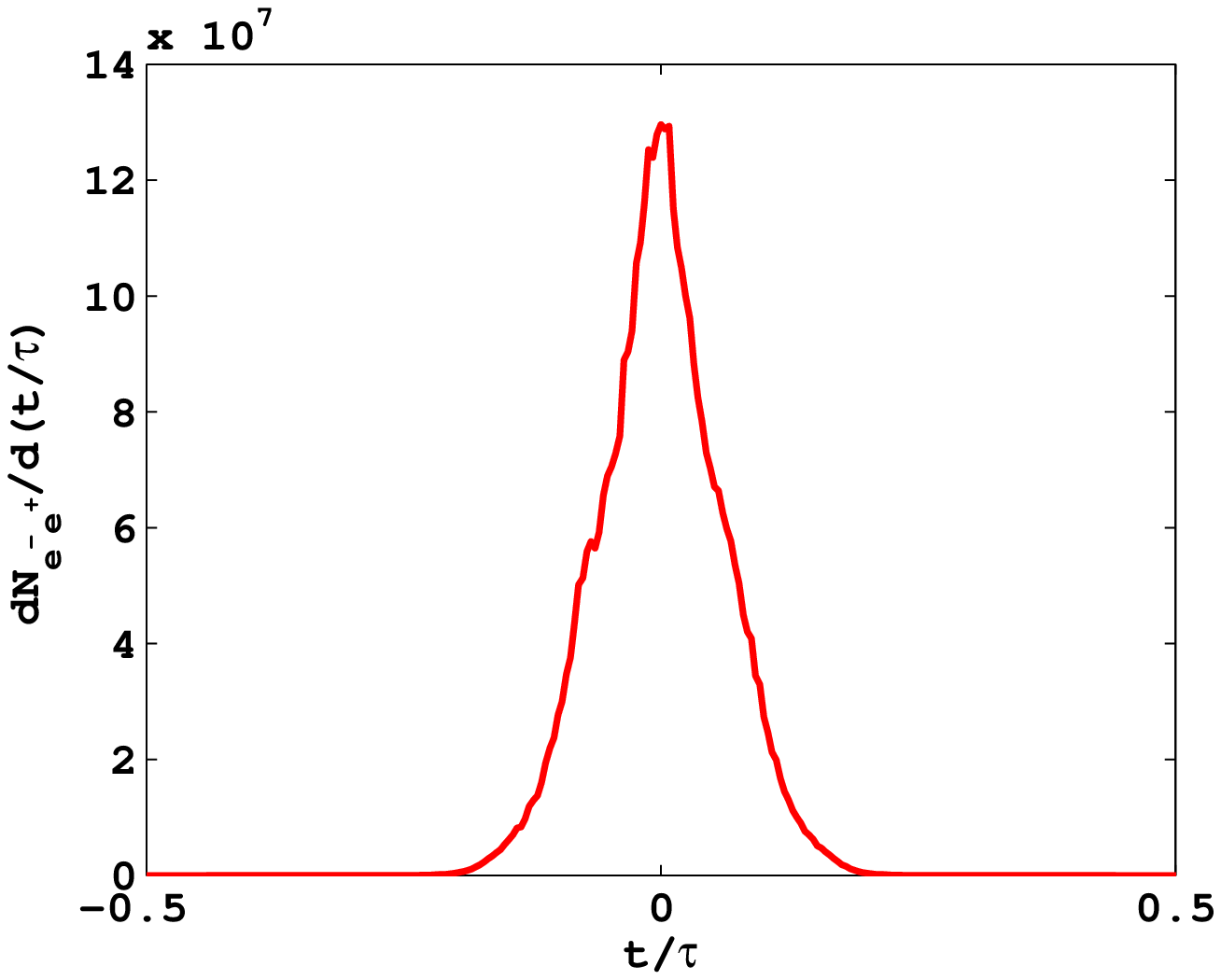}
 \end{center}
 \begin{center}
 \caption{Distribution of $dN_{e^+e^-}/d{(t/\tau)}$ as a function of  $t/\tau$ for  two counterpropagating focused laser beams with $\mu = -1$.  $E_0 = 0.0565E_S$, $\Delta = 0.1$, and $\tau=10fs$.}\label{fig:dN_vs_t_mu1_8May2016}
 \end{center}
 \end{figure}

 It is then natural to examine if one can use the expressions for the magnitude of  electric and magnetic fields in the laboratory frame instead of those for  $\epsilon$ and $\eta$ for calculating number of pairs in Eq.(\ref{N_e_average}) for the counterpropagating laser beams with $\mu = \mp 1$.
  \begin{table}[ht]
  \caption{$N_{e^+e^-}$ for $\mu = -1$ using ($|Re\textbf{E}|$,$|Re\textbf{H}|$) and ($\epsilon$,$\eta$). Here $\Delta = 0.1$, $\tau = 10fs$, and $\lambda = 1\mu m$.} 
  \centering  
  \begin{tabular}{c c c } 
  \hline\hline                        
  $I\times 10^{27}{W/cm^2}$ & $N_{e^+e^-}$($|Re\textbf{E}|$, $|Re\textbf{H}|$) & $N_{e^+e^-}$($\epsilon$, $\eta$)  \\ [0.5ex] 
  \hline                  
  $0.2$ & $3.7157$ & $3.5269$  \\ 
  \hline
  $0.3$ & $2.1308 (4)$ & $2.0135(4)$  \\
  \hline
  $0.4$ & $4.1661(6)$ & $3.9253(6)$ \\
  \hline
  $0.5$ & $1.5907(8)$ & $1.4944(8)$  \\
  \hline
  $0.6$ & $2.4276(9)$ & $2.2782(9)$  \\
  \hline
  $0.7$ & $2.0694(10)$ & $1.9375(10)$  \\
  \hline
  $0.8$ & $1.1857(11)$ & $1.1091(11)$  \\
  \hline
  $0.9$ & $5.158(11)$ & $4.82(11)$  \\
  \hline
  $1$ & $1.7912(12)$ & $1.6723(12)$ \\ [1ex]      
  \hline\hline 
  \end{tabular}
  \footnote{The numbers in the brackets indicate in powers of $10$.}
  \label{table:N_mu_1_epsilon_ReE} 
  \end{table}
 We see that the expressions of the EM fields in both the frame are same
 in the small $\xi$ approximation. So we use this field expressions and
 calculate the number of pairs in both cases which is tabulated in Table
 \ref{table:N_mu_1_epsilon_ReE}. The Table \ref{table:N_mu_1_epsilon_ReE}
 shows the numbers of pairs for $\mu = -1$, using fields
  ($|Re\textbf{E}|$, $|Re\textbf{H}|$) in the place of ($\epsilon$,
   $\eta$) in Eqn.(\ref{N_e_average}) in the 1st. column.  The second column shows the results using ($\epsilon$,
 $\eta$) in Eqn.(\ref{N_e_average}). It is seen that 
 the number of pairs are almost same in column 1 and 2.  One immediate ramification of this observation is that  one can work in the laboratory frame for colliding pulses which circularly e- or h-polarized to study the pair production. This would offer enormous simplification for analytical calculation and thus may help in getting physical insight of the underlying process.  \\
 Having discussed the structure of the electromagnetic fields in the focal region and their relationship with the reduced field invariants, we now investigate the spatio-temporal distribution of the created pairs in the focal region. For convenience we define the differential particle distribution with respect to a particular space/time coordinate when $w_{e^-e^+}$ given by Eq. (\ref{w_e_average}) is integrated over all the other coordinates except for the coordinate under the consideration. This obviously gives the derivative of $N_{e^+e^-}$ with respect to that coordinate.  Such a differential particle distribution in spatio-temporal coordinates provides a measure  to know the space-time extension of the pair production in  the focal region. The variation of $dN_{e^+e^-}/{d\xi}$ as a function of   $\xi$ for $\mu = \pm 1$ is shown in Figs. (\ref{fig:dN_vs_xi_mu1_7May2016}-\ref{fig:dN_vs_xi_mu_1_7May2016}). In both the cases, it vanishes for $\xi=0$, starts increasing  with increase in  $\xi$ for small values of $\xi$, attains a maximum value and thereafter decays exponentially with further increase in $\xi$.  The extent  of $dN_{e^+e^-}/{d\xi}$ in the transverse direction $\xi$ can be quantified by the full-width-half-maxima (FWHM) of the respective curves. FWHM is $0.144R$ for $\mu = -1$ and $0.149R$ for $\mu = 1$, where $R$ is the focal radius defined earlier. 
The differential particle distribution $dN_{e^+e^-}/{d\chi}$ as function of $\xi$  for $\mu = \pm 1$ shows spiky behaviour in the subwavelength extension. For $\mu = -1$, as seen in the Fig.(\ref{fig:dN_vs_xi_mu_1_7May2016}) the distribution possesses a prominent  peak in  the central region. There are two small but finite bumps on either side of the central peak. In the Fig.(\ref{fig:dN_chi_mu1_9May2016})  $dN_{e^+e^-}/{d\chi}$ as a function of  $\chi$ shows two peaks located at $\chi =\mp 0.0164$ for $\mu = 1$. The effective longitudinal scale length at which maximum number of particles are created is of the order of $0.0048L \approx 0.076 \mu m$ for $\mu = -1$ and $0.0732\mu m$ for $\mu = 1$.
\begin{figure}[h]
\begin{center}
\includegraphics[width=75mm]{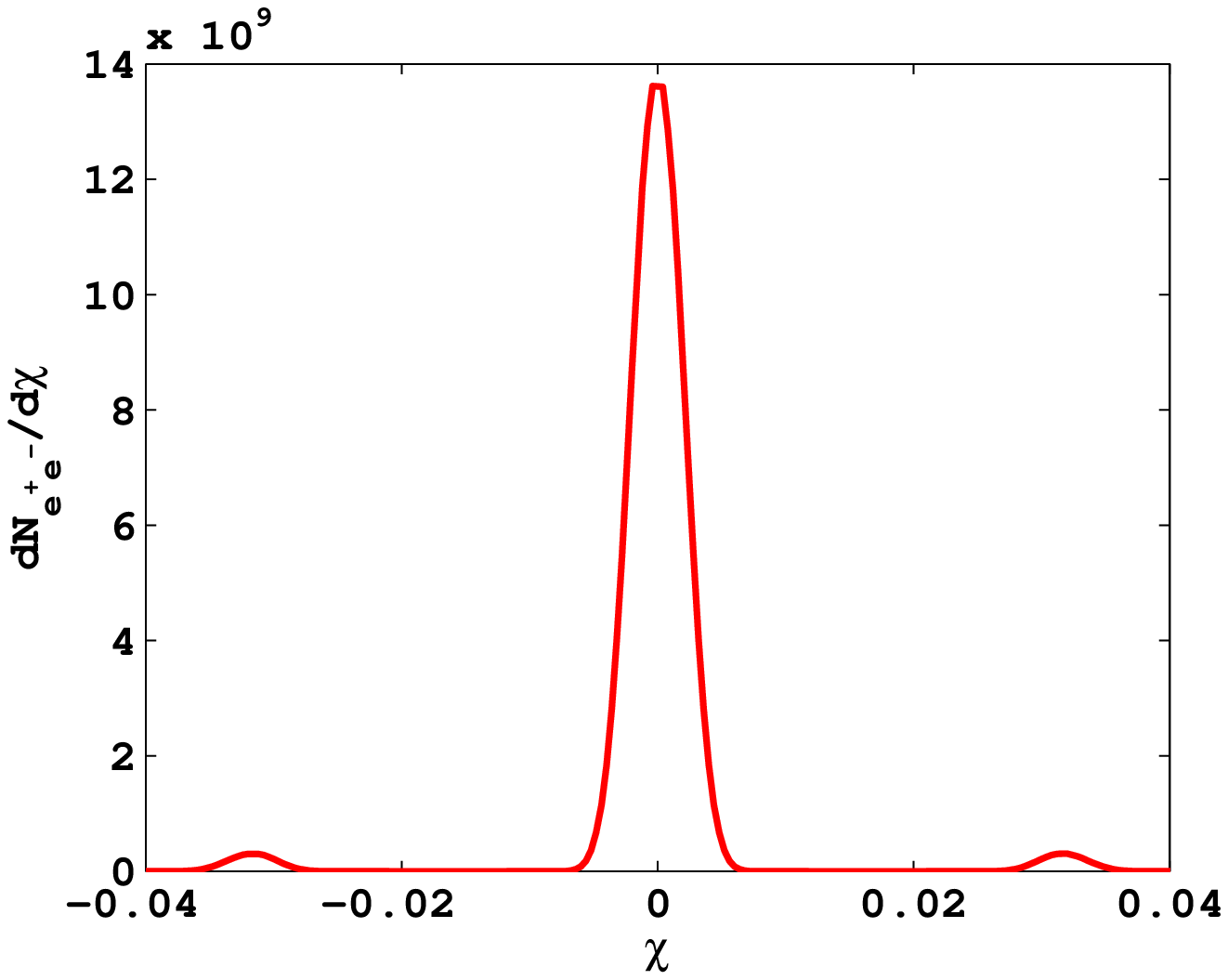}
\end{center}
\begin{center}
\caption{Distribution of $dN_{e^+e^-}/{d\chi}$ as function of $\chi$ for two counterpropagating focused laser beams having $\mu = -1$. $E_0 = 0.0565E_S$, $\Delta = 0.1$, and $\tau=10fs$.}\label{fig:dN_chi_mu_1_9May2016}
\end{center}
\end{figure}
 The differential particle distribution $dN_{e^+e^-}/{d\phi}$ as a function of  $\phi$ for $\mu = \mp 1$ is  shown in Figs. (\ref{fig:dN_vs_phi_mu_1_7May2016}-\ref{fig:dN_vs_phi_mu1_7May2016}). These figures  show  that pair production process does not have the azimuthal symmetry. It has a tendency to produce maximum number of pairs at $\phi = \pi/2$ and $3\pi/2$. FWHM for this distribution  is of the order of $0.2679 \pi$  for $\mu = -1$ and $0.2517 \pi$ for $\mu = 1$. 
The distribution of $dN_{e^+e^-}/d{(t/\tau)}$ as a function of  $t/\tau$ for $\mu = \mp 1$, is shown in Figs.(\ref{fig:dN_vs_t_mu_1_8May2016}-\ref{fig:dN_vs_t_mu1_8May2016}) which present the differential particle production with time. One can see that the particles are produced over a much shorter time duration compared to the pulse duration of the laser pulses. It is possible to estimate the  bunch duration of electrons/positrons by  FWHM of the respective curves. FWHM is  of the order of $1.4 fs$ for $\mu = -1$ and $1.351 fs$ for $\mu = 1$. 
Thus it is found that the pairs are produced over a very limited region of the focal region and over a very short time duration (compared to the pulse duration of the laser beam). The differential particle distribution as a function of  $\chi$ shows a spiky structure. For $\mu = 1$ it shows spike which are located at $\chi = \mp 0.0164$ where $\epsilon$ have peak values. The distribution of the pairs convey that central region is effective for $\mu = -1$ whereas for $\mu = 1$ the two side peak regions are important. The contribution of the other lobes in the longitudinal direction are not significant.\\
\subsection{Fields and particles for $\mu = 0$ beam:}
\begin{figure}[h]
 \begin{center}
 \includegraphics[width=75mm]{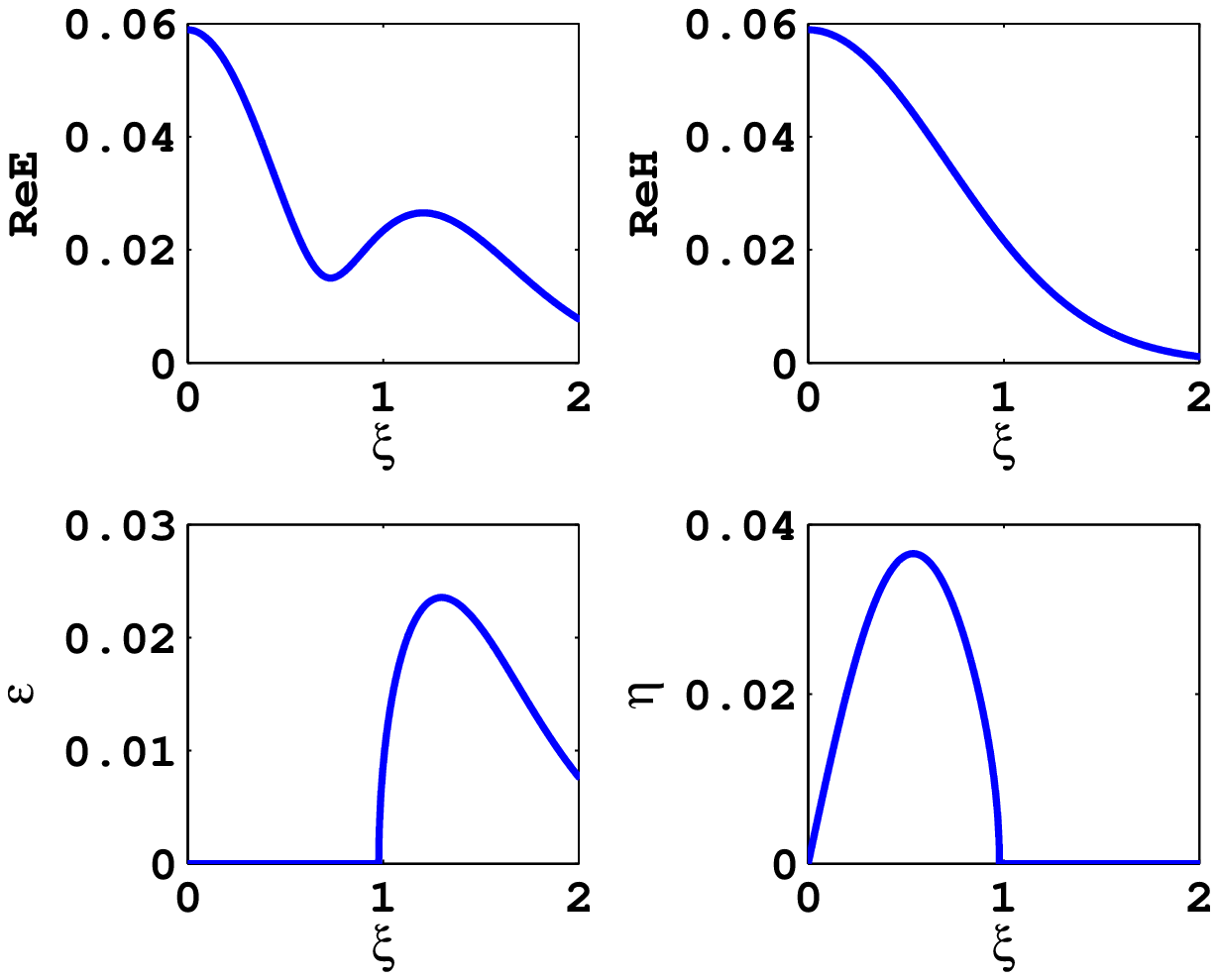}
 \end{center}
\begin{center}
\caption{Variation of $|Re\textbf{E}|$, $|Re\textbf{H}|$, $\epsilon$, and $\eta$ with dimensionless transverse variable $\xi$ of the two counterpropagating focused laser beam for $\mu =0$. Fields are normalized with $E_S$, $E_0 = 0.0565E_S$, $\chi = 0$, $\phi = 0$ and $t=0$.}\label{fig:E_H_epsilon_eta_xi_mu0_chi0}
\end{center}
\end{figure}

\begin{figure}[h]
 \begin{center}
 \includegraphics[width=75mm]{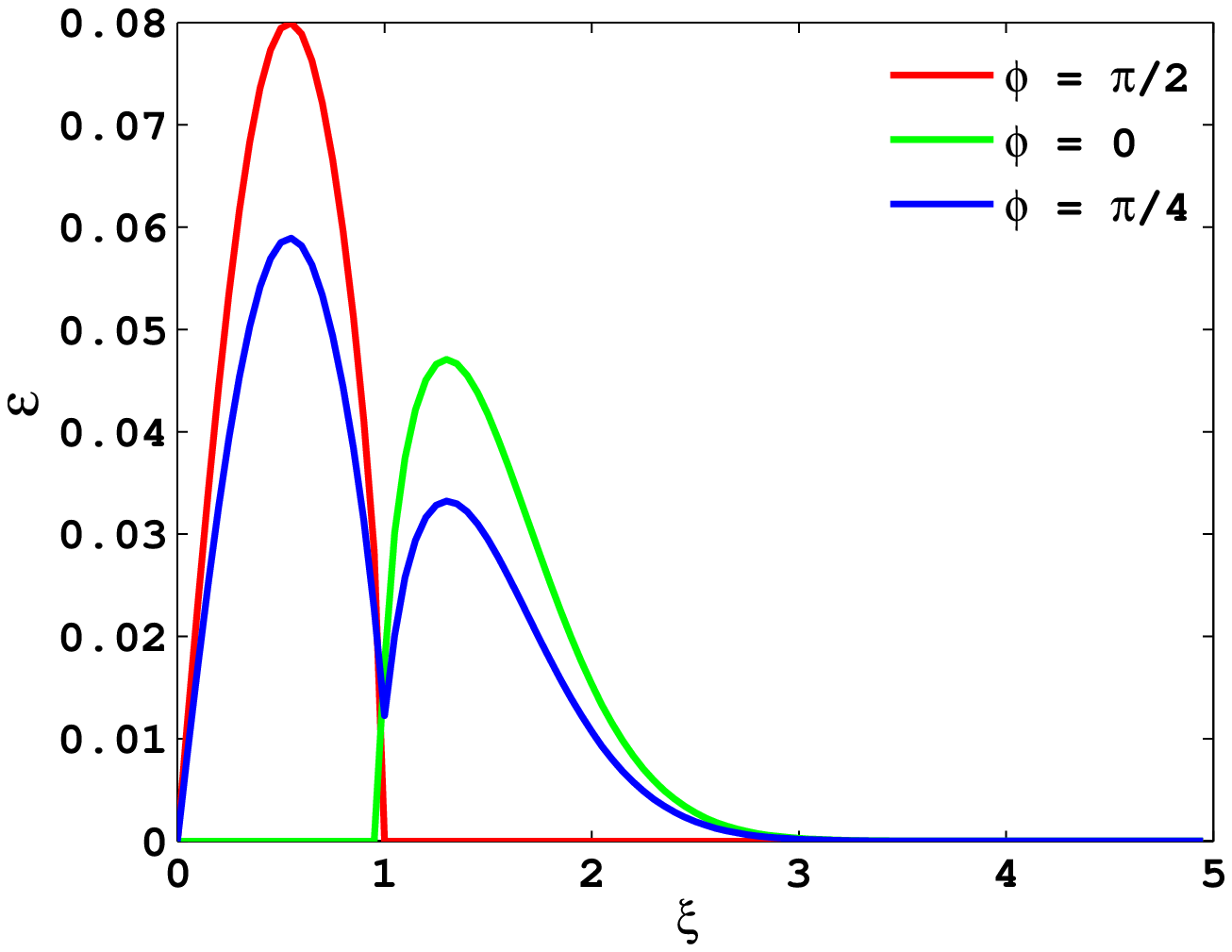}
 \end{center}
\begin{center}
\caption{Variation of $\epsilon$ with dimensionless transverse variable $\xi$ of the two counterpropagating focused laser beam of $\mu =0$ for $\phi = 0$, $\pi/2$, and $\pi/4$. Fields are normalized with $E_S$,$E_0 = 0.0565E_S$, $\chi = 0$, and $t=0$.}\label{fig:epsilon_xi_phi0}
\end{center}
\end{figure}

\begin{figure}[h]
 \begin{center}
 \includegraphics[width=75mm]{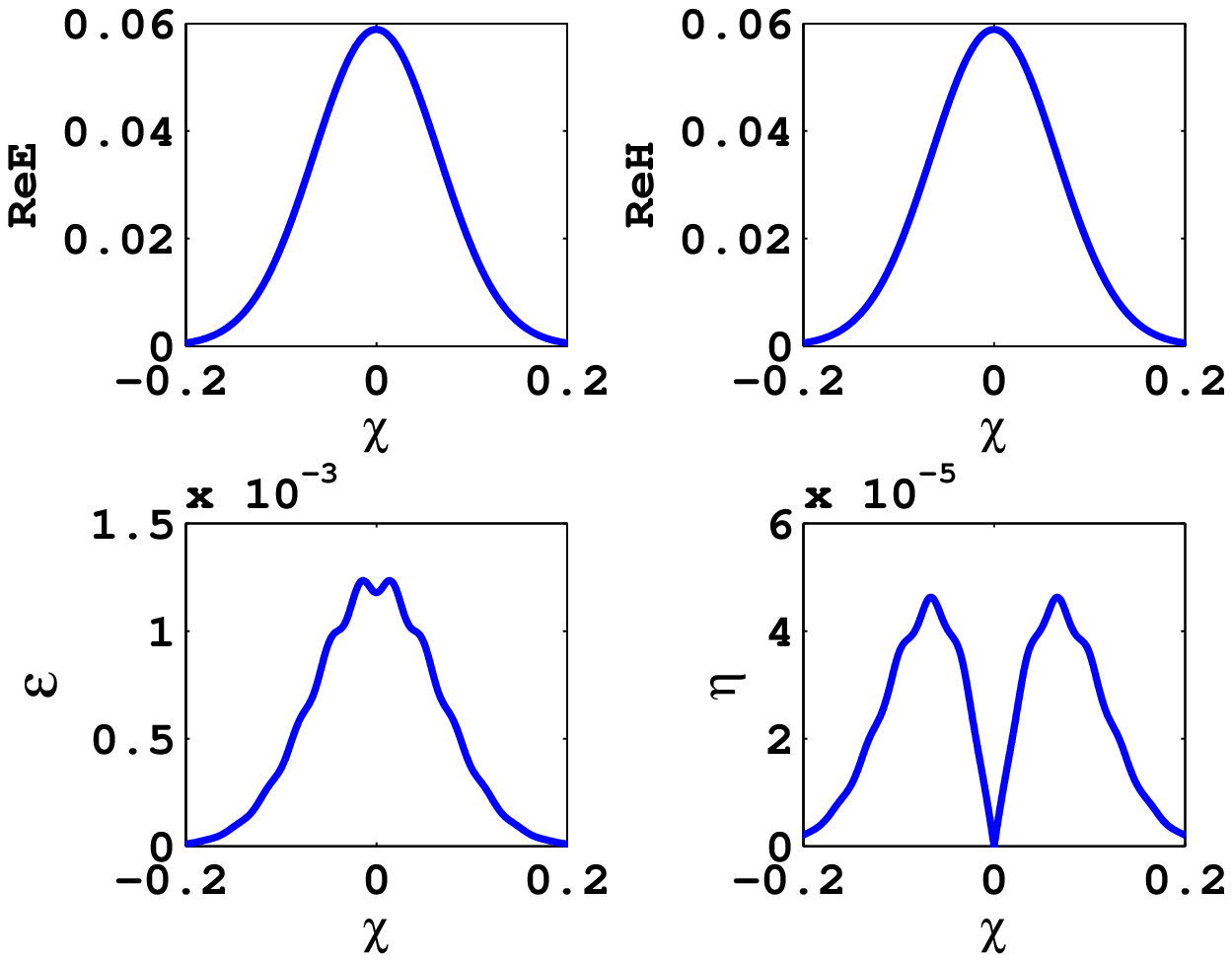}
 \end{center}
 \begin{center}
 \caption{Variation of $|Re\textbf{E}|$, $|Re\textbf{H}|$, $\epsilon$, and $\eta$ with dimensionless longitudinal variable $\chi$ of the two counterpropagating focused laser beam for $\mu = 0$. Fields are normalized with $E_S$, $E_0 = 0.0565E_S$, $\xi = 0.01$, and $t=0$.}\label{fig:E_H_epsilon_eta_chi_mu0_xi01}
 \end{center}
 \end{figure}

\begin{figure}[h]
 \begin{center}
 \includegraphics[width=75mm]{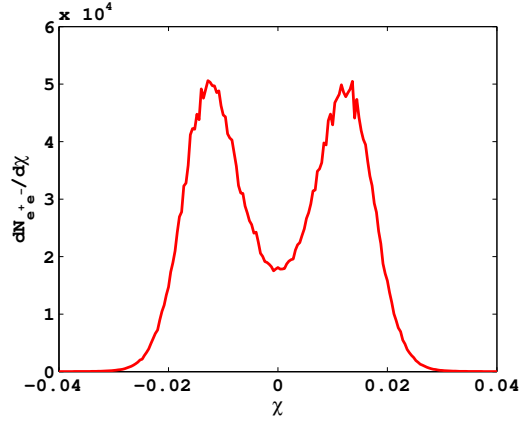}
 \end{center}
 \begin{center}
 \caption{Distribution of $dN_{e^+e^-}/{d\chi}$  as a function of  $\chi$ for two counterpropagating focused laser beams having $\mu = 0$. Here, integration has been performed over $\xi$, $\phi$,$t$ and $\chi$ is varied as a parameter. The $E_0 = 0.0565E_S$, $\Delta = 0.1$, and $\tau=10fs$.}\label{fig:dN_chi_mu0_9May2016}
 \end{center}
 \end{figure}

\begin{figure}[h]
 \begin{center}
 \includegraphics[width=75mm]{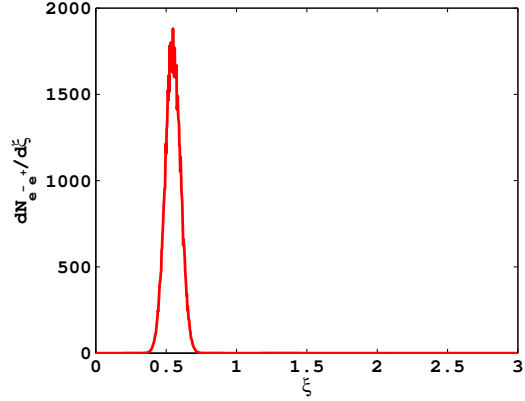}
 \end{center}
 \begin{center}
 \caption{$dN_{e^+e^-}/{d\xi}$ as a function of  $\xi$ for two counterpropagating focused laser beam having $\mu =0$. Here, integration has been performed over $\xi$, $\phi$,$t$ and $\chi$ is varied as a parameter. The value of $E_0 = 0.0565E_S$, $\Delta = 0.1$, and $\tau=10fs$.}\label{fig:dN_vs_xi_mu07May2016}
 \end{center}
 \end{figure}

\begin{figure}[h]
   \begin{center}
   \includegraphics[width=75mm]{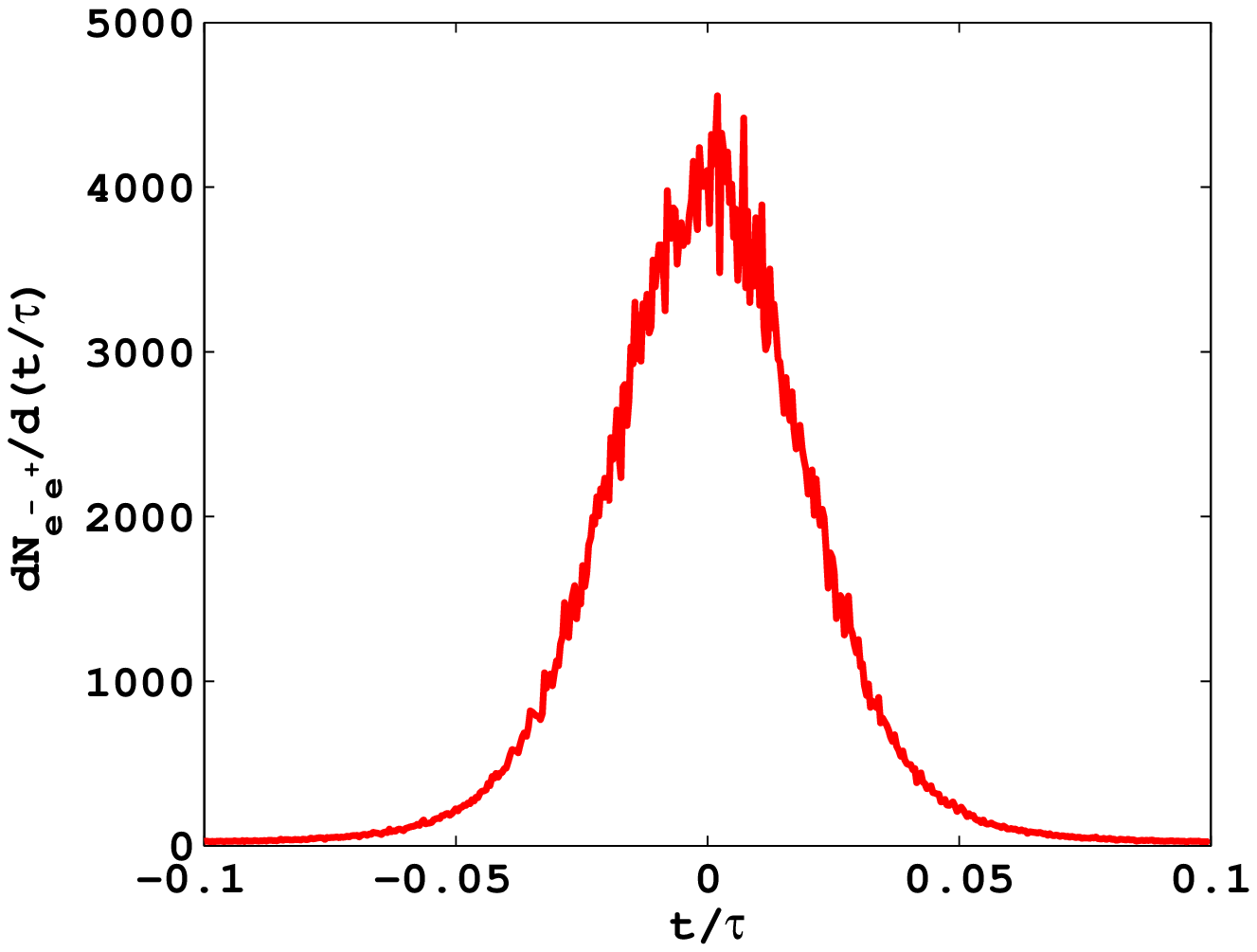}
   \end{center}
   \begin{center}
   \caption{Distribution of $dN_{e^+e^-}/{t/\tau}$ as a function of  $t/\tau$ for two counterpropagating focused laser beams having $\mu =0$. Here, integration has been performed over $\xi$, $\phi$, $\chi$ and $t$ is varied as a parameter. The value of $E_0 = 0.0565E_S$, $\Delta = 0.1$, and $\tau=10fs$.}\label{fig:dN_vs_t_mu08May2016}
   \end{center}
   \end{figure}

 Here we present the distribution of the fields in transverse and longitudinal spatial variables, $\xi$, and $\chi$ and consequently discuss the particle production mechanism.
 Figure \ref{fig:E_H_epsilon_eta_xi_mu0_chi0} shows the fields and invariants with $\xi$ for $\mu = 0$. The overall peak value of the field intensity is several order less in compare to the $\mu = \mp 1$ cases. The electric field is $\pi/2$ out of phase with the magnetic field which is also seen in the analytical expression in Eq.(\ref{mu =0 field}).
In the Fig.(\ref{fig:epsilon_xi_phi0}) we present the reduced electric field $\epsilon$ with $\xi$ for $\phi = 0$, $\pi/2$, $\pi/4$. It shows very strong dependence on $\phi$ which is basically confirmation with the analytical expression of $\epsilon$ given in Eq.(\ref{eqn:epsilon_mu0}). 
The distributions of the fields and invariants are shown in Figure \ref{fig:E_H_epsilon_eta_chi_mu0_xi01} with $\chi$ for the $\mu = 0$ beam configuration. Here the resultant field distributions do not have any interference pattern and peak value of the reduced field are several order less.
 Hence the distinguishing property of the field distribution is that the electric and magnetic fields  in  the two frames are quite different - both qualitatively and quantitatively . This non-parallelism of the fields in the lab frame can be visualised by the analytical expression of the cross product of $Re\textbf{E}$ and $Re\textbf{H}$. We calculate the $x$, $y$, and $z$-components of $Re\textbf{E}\times Re\textbf{H}$. However the $x$, $y$-components are negligible small because of the presence of the factor $\xi \Delta$ in their expressions. The most significant feature is contained  in the $z$-component of the cross-product $\textbf{C}$, which takes the form
 \begin{equation}
 C_z = \frac{4E_0^2g^2e^{-\frac{2\xi^2}{1+4\chi^2}}}{(1+4\chi^2)^{2}}[1-\frac{2\xi^2}{(1+4\chi^2)^{1/2}}\cos{\psi}].
 \end{equation}
 The above expression shows that the electric and magnetic fields are almost orthogonal to each other in the lab frame and the parallel portion of the fields goes as $\xi^2$.  
 The reduced field invariants are, therefore,   for $\mu = 0$  are much less compared to  the fields in the lab frame. This feature  of the field tells us that there is finite EM field energy flowing out of the focal region. Hence this field configuration is not efficient for the pair production.  We have already  seen that in  the expressions of $\epsilon$ and $\eta$  the leading order terms are $\xi$ dependent. They are shown in Figs.(\ref{fig:E_H_epsilon_eta_xi_mu0_chi0}-\ref{fig:epsilon_xi_phi0}) as a function of $\xi$ for different values of $\phi$ and also as a function $\chi$. One can see that the reduced field invariants are very sensitive to the  azimuthal angle Fig.(\ref{fig:epsilon_xi_phi0}).\\  
 The differential particle distribution a function of  $\chi$  shows  significantly smaller peaks  as shown in Fig. (\ref{fig:dN_chi_mu0_9May2016}). The central region shows a dip and there are two peaks on its either side. The extent of the effective region of pairs has increased in comparison  to those of  $\mu = \mp 1$ cases.  FWHM of each peak is $0.0139L$.
 Fig. (\ref{fig:dN_vs_xi_mu07May2016}) depicts the differential particle distribution as a function of  $\xi$. It shows the off-centred peak of the particle generation. The the transverse extent  of the particle distribution given by its  FWHM is $0.1269R$. 
 \begin{figure}[h]
  \begin{center}
  \includegraphics[width=75mm]{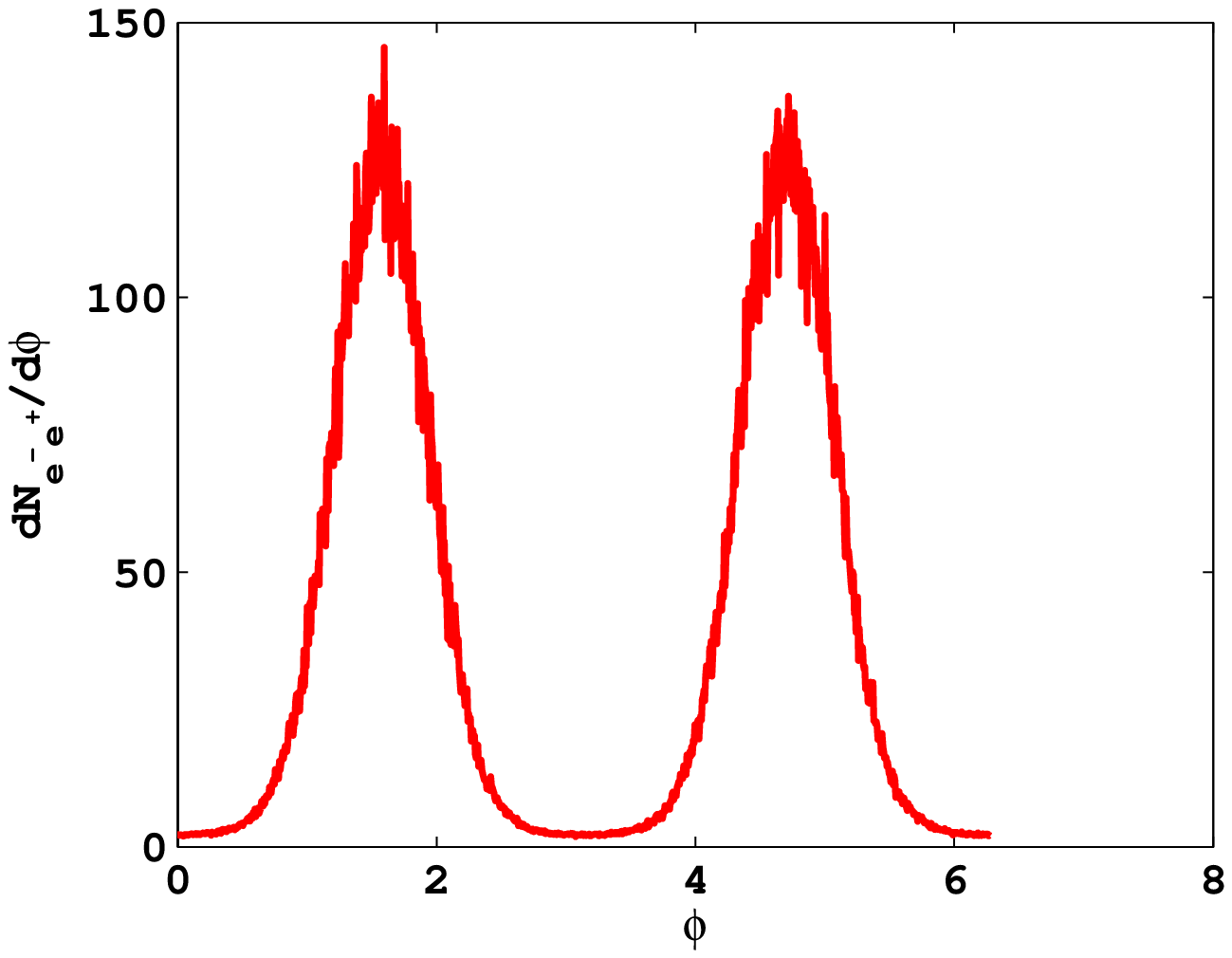}
  \end{center}
  \begin{center}
  \caption{Distribution of $dN_{e^+e^-}/{d\phi}$ as a function of  $\phi$ for two counterpropagating focused laser beams having $\mu =0$. Here, integration has been performed over $\xi$, $\chi$,  $t$ and $\phi$ is varied as a parameter. The value of $E_0 = 0.0565E_S$, $\Delta = 0.1$, and $\tau=10fs$.}\label{fig:dN_vs_phi_mu07May2016}
  \end{center}
  \end{figure} 
  In the Fig.( \ref{fig:dN_vs_phi_mu07May2016}), we present the differential particle distribution as a function of  $\phi$ for $\mu = 0$. It shows a strong dependence on $\phi$ which is directly manifested by the reduced fields distribution $\epsilon$ and $\eta$. The peaks are located at $\phi = \pi/2$ and $3\pi/2$ having FWHM of the order of $0.2782$. An interesting feature is observes in Fig. (\ref{fig:dN_vs_t_mu08May2016}) which presents the differential particle distribution as a function of time. The distribution shows a very sharp peak of FWHM $449 as$. This implies that it is possible to generate ultra short duration  particle bunches using this configuration - much shorter than what can be obtained using laser pulses with $\mu = \mp 1$.   

\section{Conclusion}
We have discussed the particle production mechanism via Schwinger mechanism in the superposition of two counterpropagating focused laser beams for $\mu = \mp 1$, and $0$. The complete features of the pairs generations have been explained on the basis of the structure of the electromagnetic fields and their relationship with the invariants and the reduced field invariants. Analytical expressions of the resultant field distribution in both the frames are discussed. These analytical expressions are used to pinpoint why colliding beam configurations with  $\mu = \mp 1$ are particularly efficient for pair production and why that corresponding to  $\mu = 0$ gives much lower number of the pairs. It has been established that the configurations with  $\mu = \mp 1$ yields  electric and magnetic fields which are almost parallel to each other in the focal region. This minimizes the energy flowing out of the focal region and thereby producing maximum number of pairs. Just opposite situation arises for the configuration $\mu = 0$. In this case the resulting electric and magnetic fields are nearly orthogonal to each other and the most of electromagnetic energy flows out of the focal region thereby effecting less number of pairs. While  $\mu = 0 $ configuration is not efficient for pair production, it offers the possibility for generating ultra short bunches of electrons/positrons. The kinematic property of the particles or the momentum distribution \cite{Popov200283} is not discussed here.  There are interesting field models such as 'e -dipole' pulse \cite{PhysRevA.86.2012Gonoskov,PhysRevLett.111.060404Gonoskov} or tightly focused $\Delta (> 1)$ fields \cite{fedotov2009electron} which are quite promising for QED processes. It would be worthwhile extending the analyses discussed here to these models. Some of the outstanding issues will be addressed in future publications..
\section{Acknowledgements}
It is a pleasure to acknowledge helpful discussions with Dr. P. A. Naik and Dr. S. Krishnagopal. 
\bibliographystyle{apsrev}
\bibliography{ref}
\end{document}